%% file: hmvcorr.tex
\documentstyle[11pt,psfig]{article}

\input{defn}

\newcommand{\axpr}{KT3}
\newcommand{\axnl}{KT4}
\newcommand{\axprsync}{KT2}

\newcommand{\treep}{\mbox{\it tree}}
\renewcommand{\cl}{\mbox{\it cl}}

\newcommand{\And}{\wedge}
\newcommand{\Or}{\vee}
\newcommand{\Imp}{\Rightarrow}
\newcommand{\Iff}{\Leftrightarrow}
\newcommand{\prove}{\vdash}

\newcommand{\bigor}{\bigvee}
\newcommand{\bigand}{\bigwedge}

\newcommand{\vp}{\varphi}
\newcommand{\vpo}{\varphi_{1}}
\newcommand{\vptw}{\varphi_{2}}
\newcommand{\vpth}{\varphi_{3}}
\newcommand{\vpn}{\varphi_{n}}

\newcommand{\alw}{\Box}

\newcommand{\ki}{K_{i}}

\newcommand{\peqi}{\approx_{i}}

\newcommand{\eqi}{\sim_{i}}

\newcommand{\runs}{{\cal R}}
\newcommand{\sysI}{{\cal I}}

\newcommand{\step}{\rightarrow}

\newcommand{\vpW}{\varphi_{W}}
\newcommand{\vpX}{\varphi_{X}}
\newcommand{\vpY}{\varphi_{Y}}
\newcommand{\forces}{\mbox{$~\|\hspace{-3pt}-~$}}

\newcommand{\vpT}{\varphi_{T}}

\newcommand{\ksi}{\Phi_{s,i}}
\newcommand{\kti}{\Phi_{t,i}}

\newcommand{\vps}{\varphi_{s}}
\newcommand{\vptt}{\varphi_{t}}

\newcommand{\vpv}{\varphi_{v}}

\newcommand{\shi}{\sigma\# i}

\newcommand{\citeyear}{\cite}
\newcommand{\KL}{\mbox{{\it KL\/}}}
\newcommand{\CKL}{\mbox{{\it CKL\/}}}
\newcommand{\ad}{\mbox{{\it ad\/}}}
\newcommand{\ex}{\mbox{{\it ex\/}}}
\newcommand{\uis}{{\it uis\/}}
\newcommand{\nl}{{\it nl\/}}
\newcommand{\nf}{{\it pr\/}}
\newcommand{\pr}{{\it pr\/}}
\newcommand{\sync}{{\it sync\/}}

\begin{document}
\begin{titlepage}
\title{Complete Axiomatizations for Reasoning About Knowledge and Time%
\footnote{This paper incorporates results from \cite{HV}, \cite{HV3}, and
\cite{Mey93}.}
}
\author{Joseph Y.\ Halpern%
\thanks{Much of the work on this paper was carried out while this author
was at the IBM 
Almaden Research Center.  IBM's support is gratefully acknowledged.  The
work was also supported in part by the NSF, under grants IRI-95-03109
and IRI-96-25901, and the Air Force Office of Scientific Research
(AFSC), under grant F94620-96-1-0323.}\\
Dept. Computer Science\\
Cornell University\\
Ithaca, NY 14853\\
halpern@cs.cornell.edu\\
http://www.cs.cornell.edu/home/halpern\\
\and
Ron van der Meyden\\
 Computing Science\\
 University of Technology, Sydney\\
 P.O. Box 123, Broadway NSW 2007\\
 Australia\\
 ron@socs.uts.edu.au\\
\and
Moshe Y. Vardi\\
 Computer Science Department\\
 Rice University\\
 Houston, TX 77005-1892\\
 vardi@cs.rice.edu\\
 http://www.cs.rice.edu/$^\sim$vardi}
\date{\today}
\maketitle
\thispagestyle{empty}
\begin{abstract}
Sound and complete axiomatizations are provided for a number of
different logics involving modalities for knowledge and time.  These
logics arise from different choices for 
various
parameters.  All
the logics considered involve the discrete time linear temporal logic
operators `next' and `until' and an operator for the knowledge of each
of a number of agents. Both the single agent and multiple agent cases
are studied: in some instances of the latter there is also an operator
for the common knowledge of the group of all agents.  Four different
semantic properties of agents are considered: whether they have a
unique initial state, whether they operate synchronously, whether they
have perfect recall, and whether they learn. The property of no
learning 
essentially dual to perfect recall.
Not all settings of
these parameters lead to recursively axiomatizable logics, but sound
and complete axiomatizations are presented for all 
the ones that do.
\end{abstract}
\end{titlepage}

\section{Introduction}

It has recently been argued that knowledge is a useful tool
for analyzing the behavior and interaction of agents in a distributed
system (see \cite{FHMV} and the references therein).
When analyzing a system in terms of knowledge, not only is the
current state of knowledge of the agents in the system relevant,
but also how that state of knowledge changes over time.  A formal
propositional logic of knowledge and time
was first proposed by Sato \citeyear{Sat}; many others have since
been proposed \cite{FHV1,Leh,LR,PR,Spaan}.  Unfortunately, while these logics
often use similar or identical notation, they differ in a number of
significant respects.

In \cite{HV2},
logics for knowledge
and time were categorized along two major dimensions: the language used
and the assumptions made on the underlying distributed system.
The properties of knowledge in a system turn out to depend in 
subtle ways on these assumptions.
The assumptions considered in \cite{HV2} concern whether agents have {\em unique
initial states}, operate {\em synchronously\/} or {\em asynchronously},
have {\em perfect recall}, and whether they
satisfy a condition called {\em no learning}.
There are 16 possible combinations of these assumptions on the 
underlying system. 
Together with 6 choices of language, this gives us 96 logics in all.
All the logics considered in the
papers mentioned above fit into the framework.  
In \cite{HV2,HV4},
the complexity of these logics is completely characterized; the results
of these papers show how the subtle
interplay of the parameters can have a tremendous impact on complexity.
The complexity results show that some of these logics cannot be
characterized axiomatically, since the  set of valid formulas for
these logics is not recursively enumerable.
Of these
96 logics, 48 involve {\em linear time} and 48
involve {\em branching time}. (The distinction between
linear and branching time essentially amounts to
whether or not we can quantify over the possible executions of a
program.)  To keep this paper to manageable length, we focus here on the
linear time logics, and provide
axiomatic characterizations of all the linear time logics
for which an axiomatization is possible at all (\ie for those logics for
which the set of valid formulas is r.e.).  

The rest of this paper is organized as follows.  In the next section, we
provide formal definitions for the logics we consider.
In Section ~\ref{background}, we review the syntax and semantics of
all the logics of knowledge and time that we consider here.
In particular, we review the
four assumptions on the underlying system
that we axiomatize in this paper.
In Section~\ref{axioms}, we state the axioms for all the systems.
In Section~\ref{framework}, we introduce the notion of {\em enriched
systems}, which form the basis for all our completeness proofs.
In Section~\ref{proofs}, we prove soundness and completeness for
the axiom systems described in Section~\ref{axioms}.
The definition
of no learning that
we use here is
slightly different from that used in
\cite{FHMV,HV}, although they agree
in many cases of interest.  We discuss the motivation for our change
in Section~\ref{nolearningremarks}.  We conclude with some further
discussion in Section~\ref{discussion}.

\section{The Formal Model: Language and Systems}\label{background}
The material in this section is
largely taken from \cite{HV2}, and is repeated here to make this paper
self-contained.  The reader is encouraged to consult \cite{HV2} for
further details and motivation.

The logics we are considering are all propositional.  Thus, we start
out with primitive propositions  $p$,  $q$, \ldots
and we close the logics under negation and conjunction, so that if
 $\varphi$  and  $\psi$  are formulas, so are  $\neg \varphi$  and
 $ \varphi \land \psi$.
In addition, we close off under modalities for knowledge and time,
as discussed below.  As usual,
we view  $\true$  as an abbreviation for  $\neg (p \land \neg p)$,
 $ \varphi \lor \psi$  as an abbreviation for  $\neg(\neg \varphi \land
\neg \psi )$, and
 $ \varphi \rimp \psi$  as an abbreviation for  $\neg  \varphi
\lor \psi$.

If we have  $m$  agents
(in distributed systems applications, this would mean a system with
$m$  processors), we add the modalities  $K_1$, \ldots,  $K_m$.
Thus, if  $\varphi$  is a formula, so is  $K_i \varphi$  (read
``agent  $i$
knows  $\varphi$'').
We take $L_i \phi$ to be an abbreviation for $\neg K_i \neg \phi$.
In some cases we also want to talk about common
knowledge, so we add the modalities  $E$  and  $C$  into the language;
 $E \varphi$  says that everyone knows  $\varphi$, while  $C
\phi$  says  $\varphi$  is common knowledge.

There are two basic temporal modalities (sometimes called {\em
operators}
or {\em connectives}): a unary
operator  $\Circ$  and a binary operator
 $\untill$.  Thus, if  $\varphi$ and  $\psi$  are formulas, then so are
 $\Circ \varphi$  (read  ``next time $\varphi$'') and  $ \varphi
\until \psi$
(read  ``$ \varphi$  until $\psi$'').
 $\Diamond \varphi$  is an abbreviation for  $\true \until \varphi$, while
$\Box \phi$  is an
abbreviation for  $\neg \Diamond \neg \varphi$.
Intuitively,  $\Circ \varphi$  says that  $\varphi$  is true at the next
point
(one time unit later),  $ \varphi \until \psi$  says that  $\varphi$
holds until  $\psi$  does,
 $\Diamond \varphi$  says that  $\varphi$  is eventually true (either
in the present or at
some point in the future), and  $\Box \varphi$  says that  $\varphi$
is always true
(in the present and at all points in the future).
In \cite{HV2}, branching time operators are also considered, which have
quantifiers over runs.  For example, $\forall \Circ$ is a branching time
operator such that $\forall \Circ \phi$ is true when  $\Circ \varphi$
is true for all possible futures.
Since we do not consider branching time operators in
this paper, we omit the formal definition here.
We take $\CKL_m$ to be the language for $m$ agents with all the modal
operators for knowledge and linear time discussed above; $\KL_m$ is the
restricted version without the common knowledge operator.

A {\em system\/} for  $m$  agents consists of a set  $\R$  of runs,
where each run  $r \in \R$  is a function from  $\IN$  to  $L^{m+1}$,
where  $L$  is some set of {\em local states}.  There is a local state
for each agent, together with a local state for the {\em environment};
intuitively, the environment keeps track of all the relevant features of
the system not described by the agents' local states, such as messages
in transit but not yet delivered.  Thus, $r(n)$ has the
form  $\< l_e, l_1 , \ldots , l_m \>$, where $l_e$ is the
state of the environment, and $l_i$ is the local state of agent $i$,
for $i = 1, \ldots, m$; such a tuple is called
a {\em global state}.   (Formally, we could view a system as a tuple
 $(\R,L,m)$, making the  $L$  and  $m$  explicit.  We have chosen
not to do so in order to simplify notation.  The  $L$  and  $m$
should always be clear from context.)
An {\em interpreted system  $\I$  for  $m$  agents} is a tuple
 $(\R , \pi )$
where  $\R$  is a system for  $m$  agents, and
 $\pi$  maps every point $(r,n) \in \R \times \IN$
to a truth assignment  $ \pi (r,n)$  to the primitive
propositions (so that  $ \pi (r,n)(p)
 \in \{ {\bf true} , {\bf false} \}$  for each primitive proposition
 $p$).%
\footnote{Note that while we are being consistent with \cite{HV2} here,
in \cite{FHMV}, $\pi$ is taken to be a function from global states (not
points) to truth values.  Essentially, this means that in \cite{FHMV} a
more restricted class of structures is considered, where $\pi$ is forced
to be the same at any two points associated with the same global state.
Clearly our soundness results hold in the more restricted class of
structures.  It is also easy to see that our completeness results hold
in the more restricted class too.  All our completeness proofs have (or
can be easily modified to have) the property that a structure is
constructed where each point is associated with a different global
state, and thus is an instance of the more restrictive structures used
in \cite{FHMV}.}

We now give semantics to  $\CKL_{m}$  and  $\KL_{m}$.
Given an interpreted system  $\I = (\R, \pi)$,
we write  $(\I,r,n) \sat  \varphi$  if the formula  $\varphi$  is true
at (or {\em satisfied by}) the point  $(r,n)$  of interpreted system  $\I$.
We define
 $\sat$  inductively for formulas of  $\CKL_{m}$  (for
 $\KL_{m}$  we just omit the clauses involving  $C$  and  $E$).  In order
to give the semantics for formulas of the form  $K_i \varphi$, we need to
introduce one new notion.
If  $r(n) = \<
l_1 , \ldots , l_m \>$,  $r'(n') =\<  l'_1 , \ldots , l'_m
\>$, and  $l_i = l'_i$, then we say that  $r(n)$  and
 $r'(n')$  are {\em indistinguishable to agent  $i$\/} and write
 $(r,n) \: \sim_i \: (r',n')$.
Of course,  $\sim_i$ is an equivalence relation on global states
(inducing an equivalence relations on points).
 $K_i \varphi$  is defined to be true at  $(r,n)$  exactly if
 $\varphi$  is true at all the points whose associated
global state is indistinguishable to  $i$  from that of
 $(r,n)$.  We proceed as follows:
\begin{itemize}
\item
 $(\I,r,n) \sat  p$  for a primitive proposition  $p$  iff  $ \pi (r,n)(p)
= {\bf true}$
\item
 $(\I,r,n) \sat   \varphi \land \psi$  iff
 $(\I,r,n) \sat  \varphi$  and  $(\I,r,n) \sat \psi$
\item
 $(\I,r,n) \sat \neg \varphi$  iff
 $(\I,r,n)$  $ \not\sat  $  $\varphi$
\item
 $(\I,r,n) \sat  K_i \varphi$  iff  $(\I,r',n') \sat  \varphi$
for all  $(r',n')$  such that  $(r,n) \: \sim_i \: (r',n')$
\item
 $(\I,r,n) \sat  E \varphi$  iff  $(\I,r',n') \sat  K_i \varphi$  for
 $i=1,\ldots,m$
\item
 $(\I,r,n) \sat  C \varphi$  iff  $(\I,r',n') \sat E^k \varphi$, for
$k=1,2,\ldots$
(where  $E^1  \varphi = E \varphi$  and  $E^{k+1}  \varphi = E E^k \phi$)
\item
 $(\I,r,n) \sat  \Circ \varphi$  iff  $(\I,r,n+1) \sat  \varphi$
\item
 $(\I,r,n) \sat   \varphi \until \psi$  iff there is some  $n' \ge n$
 such that
 $(\I,r,n') \sat  \psi$,
and for all  $n''$  with  $n \leq n'' < n'$, we have  $(\I,r,n'') \sat
\varphi$.
\end{itemize}

There is a graphical interpretation of the semantics of
$C$  which we shall find useful in the sequel.  Fix an interpreted system
$\I$.
A point  $(r',n')$  in  $\I$  is {\em reachable\/} from a point
$(r,n)$ if there exist points  $(r_0 , n_0 ) ,
\ldots , (r_k , n_k ) $  such that  $(r,n) = (r_0 , n_0 )$,
$(r',n') = (r_k , n_k )$, and for all  $j = 0 , \ldots ,
k-1$  there exists  $i$  such that
$(r_j , n_j ) \: \sim_i \: (r_j+1 , n_j+1 ) $.
The following result is well known (and easy to check).
\lem\label{Csemantics} {\rm \cite{HM2}} $(\I,r,n) \sat  C \varphi$
iff $(\I,r',n')
\sat \varphi$  for all points  $(r',n') $  reachable from  $(r,n)$.
\elem

As usual, we define a formula  $\varphi$  to be {\em valid with respect to
a class  $\C$  of interpreted systems} iff  $(\I,r,n) \sat  \varphi$
for all interpreted systems $\I \in
\C $ and points $(r,n)$ in $\I$.
A formula $  \varphi $ is {\em satisfiable with respect to $ \C $}
iff for
some $ \I \in \C $ and some point $(r,n)$ in $\I$, we have
$ (\I,r,n) \sat   \varphi $.

We now turn our attention to formally defining the classes
of interpreted systems of interest. For some of these definitions,
it will be useful to give a number of equivalent presentations.

Perfect recall means, intuitively, that an agent's local state encodes
everything that has happened (for that agent's point of view) thus far
in the run.  To make this precise,
define {\em agent $ i $'s local-state sequence at the point $ (r,n) $}
to be the sequence $ l_0 , \ldots , l_k $ of states that agent
$ i $ takes on in run $ r $ up to 
and including
time $n$, with consecutive repetitions omitted.  
For example, if from time 0 through 4 in run $ r $ agent
$ i $ goes through the sequence $ l,l,l',l,l $ of states, its history at
$(r,4)$ 
is
just $ l,l',l $.  
Roughly speaking, agent $ i $ has perfect recall if it
``remembers'' its history.
More formally, we say that {\em agent $i$ has perfect recall\/}
(alternatively,
{\em agent $i$ does not forget\/}) in
system $\R$ if at all points $(r,n)$ and $(r',n')$ in $\R$, if
$(r,n) \sim_i (r',n')$, then $r$ has the same local-state sequence at
both $(r,n)$ and $(r',n')$.

There are a number of 
equivalent
characterizations of perfect recall.  
One 
characterization
that will prove particularly useful in the comparison with the concept 
of no learning, which we are about to define, is the following.
Let $S = (s_0, s_1, s_2, \ldots)$ and
$T = (t_0, t_1, t_2, \ldots)$ be two (finite or infinite) sequences
and let $\sim$ be a relation on the elements of $S$ and $T$.
Then we say that $S$ and $T$ are $\sim$-concordant
if there is some $k$ ($k$ may be $\infty$)
and nonempty consecutive intervals $S_1, \ldots, S_k$ of $S$
and $T_1, \ldots, T_k$ of $T$ such that for all $s \in S_j$ and
$t \in T_j$, we have $s \sim t$, for $j = 1, \ldots, k$.

\lem\label{nfdef}{\rm \cite{HV,Mey93}}
The following are equivalent.
\bi
\item[(a)] Agent $i$ has perfect recall in system $\R$.
\item[(b)] For all points $(r,n) \eqi (r',n')$ in $\R$,
$((r,0), \ldots, (r,n))$ is $\sim_i$-concordant with $((r',0), \ldots,
(r',n'))$.
\item[(c)] For all points $(r,n) \eqi (r',n')$ in $\R$,
if $n > 0$, then either $(r,n-1)\eqi (r',n')$ or there
exists a number $l<n'$ such that $(r,n-1) \eqi (r',l)$ and for all $k$
with $l<k\leq n'$ we have $(r,n)\eqi (r',k)$.
\item[(d)] For all points $(r,n) \eqi (r',n')$ in $\R$,
if $ k \leq n $, then there
exists $ k' \leq n' $ such that $ (r,k) \: \sim_i \: (r',k') $.
\ei
\elem

\prf
The implications from (a) to (b), from (b) to (c) and from (c) to (d)
are straightforward. The implication from (d) to (a) can be proved by
a straightforward induction on $ n + n' $.
\eprf

\begin{figure}
\centerline{\psfig{file=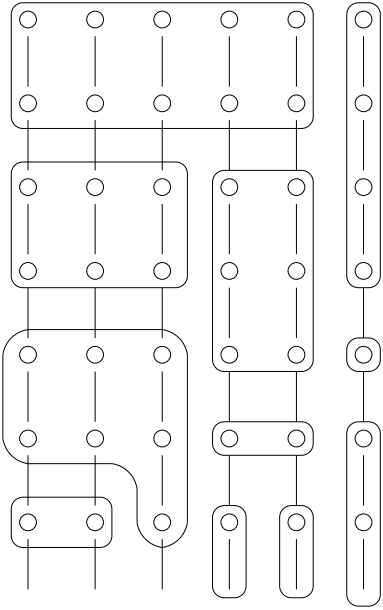,width=2in}}
\caption{A system where agent $ i $ has perfect recall}
\label{fig1}
\end{figure}

This lemma shows that perfect recall requires an unbounded number of
local states in general, since agent $ i $ may have an infinite
number of distinct histories in a given system.
A system where agent $ i $ has perfect recall is shown in
Figure~\ref{fig1},
where the vertical lines denote runs (with time 0 at the top) and
all points that $ i $ cannot distinguish are enclosed
in the same region.

We remark that the official definition of perfect recall given here is
taken from \cite{FHMV}.  In \cite{HV}, part~(d) of Lemma~\ref{nfdef}
was taken as the definition of perfect recall (which was called no
forgetting in that paper).

Roughly speaking, no learning is the dual notion to perfect recall.
Perfect recall says that if the agent considers run $r'$ possible
at the point $(r,n)$, in that there is a point $(r',n')$ that the
agent cannot distinguish from $(r,n)$, then the agent must have
considered $r'$ possible at all times in the past (\ie at all
points $(r,k)$ with $k \le n$); it is not possible
that the agent once considered $r'$ impossible and then forgot this
fact.
No learning
says that if the agent considers $r'$ possible at $(r,n)$, then
the agent will consider $r'$ possible at all times in the future;
the agent will not learn anything that will allow him to distinguish
$r$ from $r'$.
More formally, we define an
agent's {\em future local-state sequence\/} at $(r,n)$ to be the
sequence of local states $l_0, l_1, \ldots$ that the agent takes on
in run $r$, starting at $(r,n)$, with consecutive repetitions omitted.
We say agent $ i $ {\em does not learn\/} in
system $\R$ if at all points $(r,n)$ and $(r',n')$ in $\R$, if
$(r,n) \sim_i (r',n')$, then $r$ has the same future local-state
sequence at both $(r,n)$ and $(r',n')$.

Just as with perfect recall, there are a number of equivalent
formulations of no learning.
\lem\label{nldef}
The following are equivalent.
\bi
\item[(a)] Agent $i$ does not learn in system $\R$.
\item[(b)] For all points $(r,n) \eqi (r',n')$ in $\R$,
$((r,n), (r,n+1), \ldots)$ is $\sim_i$-concordant with
$((r',n'),(r',n'+1), \ldots)$.
\item[(c)] For all points $(r,n) \eqi (r',n')$ in $\R$,
either $(r,n+1)\eqi (r',n')$ or there
exists a number $l>n'$ such that $(r,n+1) \eqi (r',l)$ and for all $k$
with $l>k\ge n'$ we have $(r,n)\eqi (r',k)$.
\ei
\elem

Notice that we have no analogue to part (d) of Lemma~\ref{nfdef} in
Lemma~\ref{nldef} (where $\le$ is replaced by $\ge$).
The analogue of (d)
is strictly weaker than (a), (b), and (c), although they are equivalent
in synchronous systems (which we are about to define formally).
It was just this analogue 
of (d)
that was used to define no learning in \cite{HV,HV2}.
We examine the differences between the notions
carefully in Section~\ref{nolearningremarks}, where we provide more motivation
for the definition chosen here.

In a {\em synchronous} system, we assume that every agent has access
to a
global clock that ticks at every instant of time, and the clock reading
is part of its state.  Thus, in a synchronous system,
each agent always ``knows'' the time.
More formally,
we say that a system $\R$ is {\em synchronous\/} if for all agents
$i$ and all points $(r,n)$ and $(r',n')$,
if $ (r,n) \: \sim_i \: (r',n') $, then $ n=n' $.%
\footnote{
We remark that in \cite{HV}, a slightly weaker definition is given:
There, a system is said to be synchronous if for all runs
$ r $, if $ (r,n) \: \sim_i \: (r,n') $ then $ n=n' $.
It is easy to show (by induction on $ n $) that the two
definitions are equivalent for systems where agents have perfect recall.
In general, however, they are different.
The definition given here is the one used in \cite{FHMV,HV2}.}
Observe that in a synchronous system where $ (r,n) \: \sim_i \: (r',n) $,
an easy induction on $ n $ shows that if $ i $ has perfect recall and
$ n > 0 $, then $ (r,n-1) \: \sim_i (r',n-1) $, while if $ i $ does
not learn, then $ (r,n+1) \: \sim_i (r',n+1) $.

Finally, we say that a system $ \R $ has a {\em unique initial state}
if for all runs $ r,r' \in \R $, we have $ r(0) = r'(0) $.
Thus, if $\R $ is a system with a unique initial state, then we have
$ (r,0) \: \sim_i \: (r',0) $ for all runs $ r,r' $ in $ \R $ and all
agents $ i $.

We say that $ \I = (\R,\pi) $ is an interpreted system where agents
have perfect recall (\respc agents do not learn, time is synchronous,
there is a unique initial state) exactly if $ \R $ is a system
with that property.
We use $ {\cal C}_m $ to
denote the class of all interpreted systems for $m$ agents, and add
the %
superscripts
$ \nl $, $\nf$, $ \sync $, and $ \uis $ to denote particular subclasses
of $ {\cal C}_m $.
Thus, for example, we use ${\cal C}_m^{\nl,\nf}$ to denote the set of all
interpreted systems with $m$ agents that have perfect recall and do not
learn.  We omit the subscript $m$ when it is clear from context.

The results of \cite{HV2,HV4} 
(some of which are based on earlier results of Ladner and Reif \cite{LR})
are summarized in Table~\ref{table1}.
For $\phi \in \KL_m$,
we define  $\ad(\varphi )$  to be
the greatest number of alternations of distinct  $K_i$'s along
any branch in  $\varphi$'s parse tree.  For example,  $\ad(K_1 \neg K_2
K_1 p ) = 3$; temporal operators are not considered, so that  $\ad(K_1
\Box K_1 p ) = 1$.
(In Table~\ref{table1},
we do not consider the language $\CKL_1$.  This is
because if $m = 1$, then $C \phi$ is equivalent to $K_1
\phi$.  Thus, $\CKL_{1}$ is equivalent to $\KL_1$.)
We omit the definitions of complexity classes such as $\Pi_1^1$ and
nonelementary time $(\ex(\ad(\phi)+1,c|\phi|)$ here.  (Note that $c$
is a constant in the latter expression.)  All that matters
for our purposes is that for the cases where the complexity is $\Pi_1^1$
or co-r.e., there can be no recursive axiomatization; the validity
problem
is too hard.  We provide complete axiomatizations here for the remaining
cases.

\begin{table}[htb]
\begin{center}
\begin{tabular}{||l | l | l | l ||}
\hline
&$\CKL_m$, $m \ge 2$ & $\KL_m$, $m \ge 2$ & $\KL_1$\\
\hline
${\cal C}_m^{\nf}$,  ${\cal C}_m^{\nf,\sync}$,  ${\cal C}_m^{\nf,\uis}$,
&$\Pi_1^1$
&nonelementary time
&double-exponential\\
${\cal C}_m^{\nf,\sync,\uis}$ & &$\ex(\ad(\varphi )+1,c|\phi|)$)
&time\\
\hline
${\cal C}_m^{\nl}$,  ${\cal C}_m^{\nl,\nf}$,
${\cal C}_m^{\nl,\nf,\sync}$,
&$\Pi_1^1$
&nonelementary space
&EXPSPACE\\
${\cal C}_m^{\nl,\sync}$ & &
$\ex(\ad(\varphi ),c|\phi|)$) & \mbox{ }\\
\hline
${\cal C}_m^{\nl,\nf,\uis}$
&$\Pi_1^1$
&$\Pi_1^1$
&EXPSPACE\\
\hline
${\cal C}_m^{nl,uis}$
&co-r.e.
&co-r.e.
&EXPSPACE\\
\hline
${\cal C}_m^{\nl,\sync,\uis}$,  ${\cal C}_m^{\nl,\nf,\sync,\uis}$
&EXPSPACE
&EXPSPACE
&EXPSPACE\\
\hline
${\cal C}_m$,  ${\cal C}_m^{\sync}$,  ${\cal C}_m^{\sync,\uis}$,  ${\cal
C}_m^{\uis}$
&EXPTIME
&PSPACE
&PSPACE\\
\hline
\end{tabular}
\end{center}
\vspace{-0.3em}
\caption{The complexity of the validity problem for logics of
knowledge and time}
\label{table1}
\end{table}

\section{Axiom Systems}\label{axioms}

In this section, we describe the axioms 
and inference rules
that we need for reasoning about knowledge
and time for various classes of systems, and state the completeness
results.  The proofs of these results are deferred to
Section~\ref{proofs}.

For reasoning
about knowledge alone, the following system, with axioms K1--K5 and
rules of inference R1--R2, is well  known to be sound and complete
\cite{FHMV,Hi1}:
\begin{itemize}
\item[] K1. All tautologies of propositional logic
\item[] K2. $\ki\vp \And \ki(\vp \Imp \psi) \Imp \ki\psi$, $i = 1,
\ldots, m$

\item[] K3. $\ki\vp \Imp \vp$, $i = 1, \ldots, n$

\item[] K4. $\ki\vp \Imp \ki \ki\vp$, $i = 1, \ldots, m$

\item[] K5. $\neg \ki\vp \Imp \ki\neg \ki\vp$, $i = 1, \ldots, m$
\item[] R1. {F}rom $\vp$ and $\vp \Imp \psi$ infer $\psi$
\item[] R2. {F}rom $\vp$ infer $\ki\vp$, $i = 1, \ldots, m$
\end{itemize}
This axiom system is known as ${\rm S5}_{m}$.

For reasoning about the temporal operators individually, the
following system (together with K1 and R1), is well known to be
sound and complete \cite{FHMV,GPSS}:
\begin{itemize}
\item[] T1. $\Circ(\vp) \And \Circ(\vp\Imp \psi) \Imp \Circ\psi$
\item[] T2. $\Circ(\neg\vp) \Imp \neg \Circ \vp$
\item[] T3. $\vp \until \psi \Iff \psi \Or (\vp \And \Circ(\vp \until
\psi))$
\item[] RT1. {F}rom $\vp$  infer $\Circ \vp$

\item[] RT2. {F}rom $\vp' \Imp \neg \psi \And \Circ\vp'$
           infer $\vp' \Imp \neg (\vp \until \psi)$
\end{itemize}

The system containing the above
axioms and inference rules for both knowledge and time is called
${\rm S5}^{U}_{m}$.   ${\rm S5}^{U}_{m}$ is easily seen to be sound for
$\C_m$, the class of all systems for $m$ agents.
Given that there is no necessary
connection between knowledge and time in $\C_m$,
it is perhaps not surprising that ${\rm S5}^{U}_{m}$ should be complete
with respect to $\C_m$ as well.
Interestingly, even if we impose the requirements of
synchrony or uis, ${\rm S5}^{U}_{m}$ remains complete; our language is
not rich enough to capture these conditions.

\thm\label{S5U} ${\rm S5}^{U}_{m}$ is a sound and complete
axiomatization for the language $\KL_{m}$ with respect to $\C_m$,
$\C_m^{\sync}$, $\C_m^{\uis}$, and $\C_m^{\sync,\uis}$, for all $m$.
\ethm

We get the same lack of interaction between knowledge
in
the classes $\C_m$,
$\C_m^{\sync}$, $\C_m^{\uis}$, and $\C_m^{\sync,\uis}$ even when we add
common knowledge.
It is well known that the following two axioms
and inference rule characterize common knowledge \cite{FHMV,HM2}:
\begin{itemize}
\item[] C1. $E \phi \dimp  \bigwedge_{i=1}^m K_i \phi$
\item[] C2. $C \phi\rimp E(\phi \land C\phi$)
\item[] RC1. {F}rom $\phi \rimp E(\psi \land \phi)$ infer $\phi
\rimp C\psi$
\end{itemize}
Let ${\rm S5C}^U_m$ be the result of adding C1, C2, and RC1
to ${\rm S5}^U_{m}$.  We then have the following extension
of Theorem~\ref{S5U}.

\thm\label{S5CU} ${\rm S5C}^{U}_{m}$ is a sound and complete %
axiomatization for the language $\CKL_{m}$ with respect to $\C_m$,
$\C_m^{\sync}$, $\C_m^{\uis}$, and $\C_m^{\sync,\uis}$, for all $m$.
\ethm

If we restrict attention to systems with perfect recall or no learning,
then knowledge and time do interact.  We start by stating five axioms of
interest, and then discuss them.

\begin{itemize}
\item[] KT1. $\ki \alw \vp \Imp \alw \ki \vp$, $i = 1,\ldots, m$
\item[] KT2. $\ki \Circ \vp \Imp \Circ \ki \vp$, $i = 1, \ldots, m$.
\item[] KT3. $\ki\vpo \And \Circ(\ki\vptw \And \neg \ki\vpth) \Imp
           L_i  ((\ki\vpo) \until [(\ki\vptw) \until \neg
\vpth])$, $i = 1, \ldots, m$
\item[] KT4. $\ki\vpo \until \ki\vptw \Imp \ki(\ki \vpo \until
\ki\vptw)$, $i = 1, \ldots, m$.
\item[] KT5. $\Circ \ki \vp \Imp \ki \Circ \vp$, $i = 1, \ldots, m$.
\end{itemize}

Axiom KT1 was first
discussed by Ladner and Reif \cite{LR}.
Informally, this axiom states that if a proposition is known to be
always true, then it is always known to be true.  It is not hard
to show, using Lemma~\ref{nfdef}, that
axiom KT1 holds with perfect recall, that is, KT1 is valid in
$\C^{\nf}_m$. It was conjectured in an early draft of \cite{FHMV} that
the system ${\rm S5}_{m}^{U}+{\rm KT1}$, would be complete for
$\C^{\nf}_m$.  However, it was shown in \cite{Mey93}
that this conjecture was false.  To get completeness we need a stronger
axiom: KT3.

It is not hard to see that KT3 is valid in systems with perfect recall.
A formal proof is provided in Section~\ref{proofs},
but we can give some intuition here.  Suppose $(\I,r,n) \sat K_i \vpo
\land \Circ (K_i \vptw \land \neg K_i \vpth)$.  That means that
$(\I,r,n+1) \sat \neg K_i \vpth$, so there must be some point $(r',n')
\eqi (r,n+1)$ such that $(\I,r',n') \sat \neg \vpth$.  Because agent
$i$ has perfect recall, there must exist some $k' \le n'$ such that
$(r',k') \eqi (r,n)$.  It is not hard to show, using
Lemma~\ref{nfdef}(c), that $(\I,r',k') \sat K_i \vpo \until (K_i
\vptw \until \neg \vpth)$.  It follows that $(\I,r,n) \sat
L_i (K_i \vpo \until (K_i \vptw \until \neg \vpth))$.

In the presence of the other axioms, KT3 implies KT1.

\lem KT1 is provable in ${\rm S5}_{m}^{U}+{\rm KT3}$. \elem

\prf Note that by purely temporal reasoning, 
we 
can show
$\prove
\alw \vp \Iff \alw \alw \vp$. Using R2 and K2, this implies that
$\prove \ki \alw \vp \Iff \ki \alw \alw \vp$.
Now if $\vpo = \vptw = {\it true}$, then KT3
simplifies to $\Circ \neg \ki \vpth \Imp \neg \ki \alw \vpth$.  In
particular, taking the contrapositive, substituting $\vpth = \alw
\vp$, and using T2, we obtain $\prove \ki \alw \alw \vp \Imp \Circ \ki
\alw \vp$, which yields $\prove \ki \alw \vp \Imp \Circ
\ki \alw \vp$ by the equivalence noted above.
It is also straightforward to show 
that $\alw \vp \Imp \vp$, from which it follows, using K2 and R2, that 
$\prove \ki \alw
\vp \Imp \ki\vp$. The axiom KT1 now follows using the rule
RT2. \eprf

KT3 turns out to be strong enough to give us completeness, with or
without the condition uis.

\thm\label{KT3} ${\rm S5}^{U}_{m}+{\rm KT3}$ is a sound and complete
axiomatization for the language $\KL_{m}$ with respect to $\C_m^{\nf}$
and $\C_m^{\nf,\uis}$, for all $m$.
\ethm

Theorem~\ref{S5U} shows that requiring synchrony or uis does not have an
impact when we consider the class of all systems---$\C_m$, $\C^\sync_m$,
$\C^\uis_m$, and $\C^{\sync,\uis}_m$ are all axiomatized by
${\rm S5}_{m}^{U}$---and Theorem~\ref{KT3} shows that adding uis does
not have an impact in the presence of perfect recall.
However, requiring synchrony
does have an impact in the presence of perfect recall.  It is easy to see
that KT2 is valid in $\C_m^{\nf,\sync}$, and it clearly is not valid
in $\C_m^{\nf}$.  Moreover, KT2 suffices for completeness in
$\C_m^{\nf,\sync}$; we do not
need the complications of KT3.

\thm\label{KT2} ${\rm S5}^{U}_{m}+{\rm KT2}$ is a sound and complete
axiomatization for the language $\KL_{m}$ with respect to
$\C_m^{\nf,\sync}$ and $\C_m^{\nf,\sync,\uis}$, for all $m$.
\ethm

KT4 is the axiom that characterizes no learning.  More precisely, we
have
\thm\label{KT4} ${\rm S5}^{U}_{m}+{\rm KT4}$ is a sound and complete
axiomatization for the language $\KL_{m}$ with respect to
$\C_m^{\nl}$ for all $m$.
\ethm

Unlike previous cases, the uis assumption is not
innocuous in the presence of nl.  For one thing, it is not hard to
check that assuming uis leads to extra properties.
Indeed, as Table~\ref{table1} shows, if $m \ge 2$,
then assuming a unique initial state along with no learning results in a
class of systems that do not have a recursive axiomatic
characterization, since the validity problem is co-r.e.
On the other hand, if there is only one
agent in the picture, things simplify.  No learning together with uis
implies perfect recall.  Thus, we get

\thm\label{KT3+4} ${\rm S5}^{U}_{m}+{\rm KT3 + KT4}$ is a sound and
complete axiomatization for the language $\KL_{m}$ with respect to
$\C_m^{\nl,\nf}$ for all $m$.  Moreover, it is a sound and complete
axiomatization for the language $\KL_{1}$ with respect to
$\C_1^{\nl,\nf,\uis}$. \ethm

In synchronous systems with no learning,
things again become  simpler.  KT5, the converse of
KT2, suffices to characterize such systems.
\thm\label{KT5}
${\rm S5}^{U}_{m}+{\rm KT5}$ is a sound and complete
axiomatization for the language $\KL_{m}$ with respect to
$\C_m^{\nl,\sync}$.
\ethm
Of course, it follows from Theorem~\ref{KT5} that KT4 can be derived in
the system ${\rm S5}^{U}_{m}+{\rm KT5}$ (although this result takes some
work to prove directly).

Not surprisingly, if we combine perfect recall, no learning, and
synchrony, 
then
KT2 and KT5 give us a complete axiomatization.

\thm\label{KT2+5} ${\rm S5}^{U}_{m}+{\rm KT2 + KT5}$ is a sound
and complete axiomatization for the language $\KL_{m}$ with respect to
$\C_m^{\nl,\nf,\sync}$ for all $m$.
\ethm

Finally, it can be shown that when we combine no learning, synchrony,
and uis, then not only do all agents consider the same worlds possible
initially, but they consider the same worlds possible at all times.
As a result, the axiom $K_i \phi \dimp K_j \phi$ is valid in this case.
This allows us to reduce to the single-agent case.
Moreover, as we
observed above, in the single-agent case, no learning and uis imply
perfect recall.
Thus, we get the following result.

\thm\label{KT2+6} ${\rm S5}^{U}_{m}+{\rm KT2 + KT5} + \{K_i \phi \dimp
K_1 \phi\}$ is a sound
and complete axiomatization for the language $\KL_{m}$ with respect to
$\C_m^{\nl,\sync,\uis}$ and $\C_m^{\nl,\nf,\sync,\uis}$ for all $m$.
\ethm

A glance at Table~\ref{table1} shows that we have now provided
axiomatizations for all the cases where complete axiomatizations exist.
(Notice that for the language $\CKL_m$, if $m = 1$, then common
knowledge reduces to knowledge, while if $m > 1$, then
complete axiomatizations can exist
only for $\C_m$, $\C_m^{\sync}$, $\C_m^{uis}$,
$\C_m^{sync,uis}$, $\C_m^{\nl,\sync,\uis}$, and
$\C_m^{\nl,\nf,\sync,\uis}$.  The first four cases were dealt with in
Theorem~\ref{S5CU}, while in the last two, as we have observed, common
knowledge reduces to the knowledge of agent 1.)

\section{A Framework for Completeness Proofs}\label{framework}

In this section we develop a general framework for completeness proofs
that reduces the work required in each of the different completeness
results to a single lemma.

A formula $\psi$ is said to be {\em
consistent\/} in a logic $L$ if it is not the case that $\prove_L \neg
\psi$.  For each of the pairs of logic $L$ and class of systems $\C$
we consider, the proof that $L$ is complete with respect to $\C$
proceeds by constructing for every formula $\psi$ consistent with
respect to $L$, a system in $\C$ containing a point at which $\psi$
is true.
All the results in this section hold for every logic containing
${\rm S5}_m^U$, except for Lemma~\ref{Cforce}, which mentions common
knowledge.  This lemma holds for every logic containing ${\rm S5C}_m^U$.
Rather than mentioning the logic $L$ explicitly in each case, we just
write $\prove$ rather than $\prove_L$; 
the intended logic(s) will be clear from context. 
We also fix the formula $\psi$, which
is assumed to be consistent with respect to $L$.

A finite sequence $\sigma = i_{1}i_{2}\ldots i_{k}$ of agents,
possibly equal to the null sequence $\epsilon$, is called an {\em
index\/} if
$i_{l}\not = i_{l+1}$ for all $l<k$. We
write $|\sigma |$ for the length $k$ of such a sequence; the null
sequence has length equal to 0.

If $S$ is a set, and $S^{*}$ is the set of all finite sequences over
$S$, we define the absorptive concatenation function $\#$ from
$S^{*}\times S$ to $S^{*}$ as follows. Given a sequence $\sigma$ in
$S^{*}$ and an element $x$ of $S$, we take $\sigma \# x = \sigma$ if
the final element of $\sigma$ is $x$.  If the final element of
$\sigma$ is not equal to $x$ then we take $\sigma \# x$ to be $ \sigma
x$, i.e. the result of concatenating $x$ to $\sigma$.  We shall have
two distinct uses for this function, applying it to primarily to
sequences of agents, and sometimes to sequences of ``instantaneous
states'' of agents in the context of asynchronous
systems.
If $\psi \in CKL_m$,
for each $k \ge 0$,
we define the $k$-closure $cl_{k}(\psi)$,
and for each
agent $i$, we define the $k,i$-closure $cl_{k,i}(\psi)$.
The definition of these sets proceeds by mutual recursion:
First, we let the {\em basic closure\/} $cl_{0}(\psi)$ be the
smallest set containing $\psi$ that is closed under subformulas,
contains $\neg \phi$ if it contains $\phi$ and $\phi$ is not of the form
$\neg \phi'$,
contains $EC\phi$ if it contains $C \phi$, and contains $K_1
\phi,\ldots, K_n \phi$ if it contains $E\phi$.  (Of course, the last two
clauses do not apply if $\psi$ is in $KL_m$, and thus
does not mention common knowledge.)
If $i$ is a agent,
we take
$cl_{k,i}(\psi)$ to be the union of $cl_{k}(\psi)$ with the set of
formulas of the form $\ki(\vpo \Or\ldots \Or\vpn)$ or $\neg \ki(\vpo
\Or\ldots \Or\vpn)$, where the $\vp_{l}$ are distinct formulas in
$cl_{k}(\psi)$.
Finally,
$cl_{k+1}(\psi)$ is defined to be
$\union_{i=1}^m cl_{k,i}(\psi)$.

If $X$ is a finite set of formulas we write $\vpX$ for the conjunction
of the formulas in $X$. A finite set $X$ of formulas is said to be
consistent if $\vpX$ is consistent.  If $X$ is a finite set of
formulas and $\vp$ is a formula we write $X\forces \vp$ when $\prove
\vpX \Imp \vp$.  Clearly if $X\forces \vpo$ and $\prove \vpo \Imp
\vptw$ then $X\forces \vptw$.

Suppose $Cl$ is a finite set of formulas with the property that for all
$\vp\in Cl$, either $\neg \vp\in Cl$ or $\vp$ is of the form $\neg \vp'$
and $\vp'\in Cl$.  (Note that the sets $cl_{k}(\psi)$ and
$cl_{k,i}(\psi)$ have this property.)  We define an {\em atom\/}
of $Cl$ to be a maximal consistent subset of $Cl$.  Evidently, if $X$ is
an atom of $Cl$ and $\vp\in Cl$, then either $X\forces \vp$ or $X
\forces \neg \vp$. 
Thus, we have that 
\lem\label{atomiclem} $\prove \bigvee_{X\ {\rm
an\; atom\; of}\ Cl} \phi_X$. \elem

We begin the construction of the model of $\psi$ by first
constructing a {\em pre-model,} which
is a structure $\langle S, \step, \approx_{1},\ldots,\approx_{n}
\rangle$ consisting of a
set $S$ of states,
a binary relation $\step$ on $S$, and for each
agent $i$ an equivalence relation $\peqi$ on $S$.
Recall from Section~\ref{background} that for a formula $\phi \in
\KL_m$, the alternation depth $\ad(\vp)$ is the
number of alternations of distinct operators $\ki$ in $\vp$.
Let $d = \ad(\psi)$ if $\psi \in \KL_m$; otherwise (that is, if $\psi$
mentions the modal operator $C$), let $d = 0$.

The set $S$ consists of all the pairs
$(\sigma, X)$ such that $\sigma$ is an index,
$|\sigma| \le d$,
and
\begin{enumerate}
\item if $\sigma = \epsilon$ then $X$ is an atom of $cl_{d}(\psi)$, and
\item if $\sigma = \tau i$ then $X$ is an atom of $cl_{k,i}(\psi)$, where $k= d-|\sigma|$.
\end{enumerate}
The relation $\step$ is defined so that $(\sigma,X)
\step (\tau,Y)$ iff $\tau = \sigma$ and the formula $\vpX \And
\Circ \vpY$ is consistent.  If $X$ is an atom we write $X/\ki$ for the
set of formulas $\vp$ such that $\ki\vp \in X$.
We say that states $(\sigma, X)$ and $(\tau, Y)$ are
{\em $i$-adjacent\/} if $\sigma \# i= \tau\# i$.
The relation $\peqi$ is
defined so that $(\sigma,X)\peqi (\tau,Y)$ iff
$\sigma$ and $\tau$ are $i$-adjacent and $X/\ki = Y/\ki$.
Clearly, $i$-adjacency is an equivalence relation, as is
the relation $\peqi$.

A {\em $\sigma$-state (for $\psi$)\/} is a pair
$(\sigma,X)$ as above.  Thus $(\sigma,X)$ is the unique $\sigma$-state
with atom $X$.  If $s=(\sigma,X)$ is a state, we define $\vps$ to be the
formula $\vpX$, and write $s\forces \vp$ for $\prove \vps \Imp \vp$.
We say that the state $s$ {\em directly decides\/} a formula $\vp$ if
either
(a) $\vp \in X$ or (b) $\neg \vp \in X$ or (c)
$\vp = \neg \vp'$ and $\vp'\in X$. Note that this implies that either
$s\forces \vp$ or $s\forces \neg\vp$.
In case this latter condition holds we say simply that $s$ {\em decides\/}
$\vp$. Note that if $\sigma =\tau i$ then
each $\sigma$-state directly decides every formula in $cl_{d-|\sigma|,i}
(\psi)$.
Also,
every $\epsilon$-state directly decides every formula in
$cl_{d}(\sigma)$.

\lem\label{lem:levels}
If $s$ and $t$ are $i$-adjacent states, then the same formulas of
the form $\ki\phi$ are directly decided by $s$ and $t$.
\elem

\prf
Suppose that $s$ and $t$ are $i$-adjacent, $s$ is a $\sigma$-state,
$t$ is a $\tau$-state.
Clearly if $\sigma = \tau$, then
$s$ and $t$ directly decide the same formulas (and, {\em a fortiori}, the same
formulas of the form $\ki \phi$) since they are both maximal
consistent subsets of the same set of formulas.  If $\sigma \ne \tau$,
then either $\sigma = \tau i$ or $\tau = \sigma i$.  By symmetry,
it suffices to deal with the case $\sigma = \tau i$.
By definition, $s$ directly decides the $\ki$-formulas in $\cl_{d - |\sigma|,i}
(\psi)$, while $t$ directly decides the $\ki$-formulas in
$\cl_{d- |\tau|,j}(\psi)$
if $\tau = \tau' j$ or $\cl_d(\psi)$ if $\tau = \epsilon$.
We leave it to the reader to check that it follows that
the $\ki$-formulas directly decided by both $s$ and $t$ are
precisely those in $\cl_{d - |\sigma|,i}(\psi)$.
\eprf

 If $s$ is a $\sigma$-state, we take $\ksi$ to
be the disjunction of the
formulas $\vptt$, where $t$ ranges over the $\sigma$-states satisfying
$s\peqi t$, and we take
$\ksi^+$ to be the disjunction of the
formulas $\vptt$, where $t$ ranges over the $(\shi)$-states satisfying
$s\peqi t$.%
\footnote{It can be shown that if $|\shi| \le d$, then $\ksi$ is 
logically
equivalent to $\ksi^+$, but we do not need this fact here.}
Observe that because $\peqi$ is an
equivalence relation
we have that if $s\peqi t$ then $\ksi = \kti$ and $\ksi^+ = \kti^+$.
The
following result lists a number of knowledge formulas decided by states.

\lem\label{lem:kforce}
\item[(a)]
If $s$ is a $\sigma$-state and $t$ is a $\sigma$-state or
$(\shi)$-state such that
           $s \not \peqi t$, then  $s\forces \ki\neg \vptt$.
\item[(b)]
For all $\sigma$-states $s$, we have $s\forces \ki\ksi$; in addition, if
$|\sigma\#i| \le d$, then $s \forces \ki\ksi^+$.
\item[(c)] For all $\sigma$-states $s$ and $(\shi)$-states $t$ with
$s\peqi t$, we have $s\forces L_i \vptt$.
\item[(d)]
If $s$ is a $\sigma$-state and $t$ is a $(\shi)$-state
such that $s \not \peqi t$, then  $t\forces \neg \ki \ksi^+$.
\elem

\prf
For (a), suppose that $s\not \peqi t$, where
$s=(\sigma,X)$ and
$t=(\tau,Y)$, where $\tau$ is either $\sigma$ or $\sigma \#i$.
Then $X/\ki \not = Y/\ki$ so
either there exists a formula $\ki\vp\in X$ such that $\ki\vp \not \in
Y$ or there exists a formula $\ki\vp\in Y$ such that $\ki\vp \not \in
X$. 
As the states $s$ and $t$ are $i$-adjacent, by
Lemma~\ref{lem:levels}, in either case the formula $\ki \vp$ is
directly decided by both the states $s$ and $t$. In the first case, we
have that $\prove \vptt \Imp \neg \ki\vp$
and
hence, using R2, that
$\prove \ki(\ki\vp \Imp \neg \vptt)$. By K4 we obtain from the fact that
$\ki\vp\in
X$ that $s\forces \ki\ki\vp$. It now follows using K2 that $s \forces
\ki\neg \vptt$. In the second case, we
have that $\prove \vptt \Imp \ki\vp$, hence, using R2, that
$\prove \ki(\neg \ki\vp \Imp \neg \vptt)$. By K4 we obtain from the fact
that $\neg \ki\vp\in
X$ that $s\forces \ki \neg \ki\vp$. It now follows using K2 that $s
\forces \ki\neg \vptt$.

For (b), 
by Lemma~\ref{atomiclem}, we have that
 $\prove \bigvee_{X\ {\rm an\; atom\; of}\
cl_{k,i}(\psi)} \phi_X$.
Hence, by R2 we obtain that $\prove \ki\bigvee_{
\sigma-{\rm states}\ t} \vptt$.  It follows from
this using (a) and K2 that $s\forces \ki\ksi$.
If $|\sigma\#i| \le d$, then a similar argument shows that
$s\forces \ki\ksi^+$.

For (c), suppose that $s= (\sigma,X)\peqi (\sigma\# i,Y)= t$ and $k=
d-|\sigma \# i|$. We claim first that if $W=Y\cap cl_{k}(\psi)$, then
$s\forces \ki\neg \vptt \Iff \ki\neg \vpW$. This is because the fact that
$Y$ is a subset of $cl_{k,i}(\psi)$ implies that all
formulas $\vp$ in $Y\setminus W$ are of the form $\ki\vp'$ or $\neg \ki
\vp'$, hence $\vp
\in X$ if and only if $\vp \in Y$. Also, by K4 and K5 we have that
$s\forces \ki\ki\vp'$ when $\ki\vp'\in X$ and $s\forces \ki\neg
\ki\vp'$ when $\ki\vp' \not\in X$. It follows using K2 that $s\forces
\ki\vp_{Y\setminus W}$. Since $\vptt$ is equivalent to $\vp_{W} \And
\vp_{Y\setminus W}$, we obtain using K2 that $s\forces \ki\neg \vptt \Iff
\ki\neg \vpW$.

Now by K3 we have $\prove \vptt \Imp L_i \vptt$. Further, the
argument of the previous paragraph also shows $t \forces \ki \neg\vptt
\Iff \ki\neg \vpW$, so we obtain that $t\forces L_i \vpW$. 
But $\neg \vpW$ is equivalent to the disjunction of a subset
$\{ \vpo, \ldots ,\vpn\}$ of $cl_{k}(\psi)$. Let $\alpha$ be
the formula $\ki(\vpo \Or\ldots \Or \vpn)$, which is equivalent to
$\ki\neg\vpW$.  It follows from the definition of $cl_{k,i}(\psi)$
that $\alpha$ is in $cl_{k,i}(\psi)$, hence directly decided by both
$t$ and $s$.  Consequently, $\alpha$ is not in $Y$, since $t\forces
\neg \alpha$.  Because $X/\ki = Y/\ki$, the formula $\alpha$ is not in
$X$ either, so $s\forces \neg \alpha$.  Applying the fact that
$\alpha$ is equivalent to $\ki\neg\vpW$, we see that $s\forces
L_i \vpW$.  The equivalence of the previous paragraph now
yields that $s\forces L_i \vptt$.

For (d), note that if $t$ and $v$ are distinct $(\sigma\# i)$-states
then $t\forces \neg \vpv$. Thus, if $s$ is a $\sigma$-state such that
$s\not \peqi t$ then $t\forces \neg \ksi^+$, which implies, using
K3, that $t\forces \neg \ki \ksi^+$.
\eprf

If $T$ is a set of states, then
we write $\vpT$ for the disjunction of the formulas $\vptt$ for $t$ in
$T$. Using RT1, T1, and T2, the following result is immediate from the
fact
that  $s\not\step t$ implies 
$\prove \vps \Imp \neg \Circ \vptt$, together
with the fact that $\prove
\bigvee_{s\ {\rm a}\; \sigma-{\rm state}}\vps$, 
which follows from Lemma~\ref{atomiclem}.

\lem\label{lem:nforce}
Let $s$ be a state and let $T$ be the set of states $t$ such that
$s\step t$. Then $s\forces \Circ \vpT$
\elem

The next result provides a useful way to derive formulas containing the
until operator.

\lem\label{lem:until}
For all formulas $\alpha,\beta$ and $\gamma$, if
$\prove \alpha \Imp \neg \gamma$
and $\prove \alpha \Imp \Circ( \alpha \Or (\neg \beta \And \neg \gamma))$
then $\prove \alpha \Imp \neg(\beta \until \gamma)$.
\elem

\prf
Suppose that
$\prove \alpha \Imp \neg \gamma \And \Circ( \alpha \Or (\neg \beta \And
\neg \gamma))$.
By T3, we obtain that
$\prove \alpha \And (\beta \until \gamma)\Imp
  \neg \gamma \And \Circ(\beta \until \gamma)
  \And \Circ( \alpha \Or (\neg \beta \And \neg \gamma))$. Since,  by
T3 again,
$\prove \beta \until \gamma \Imp \neg(\neg \beta \And \neg \gamma)$,
it follows
using T1 and RT1 that
$\prove \alpha \And (\beta \until \gamma) \Imp
  \neg \gamma \And \Circ( \alpha \And (\beta \until \gamma))$. Now using
RT2 we obtain
$\prove \alpha \And (\beta \until \gamma) \Imp \neg (\beta \until \gamma)$,
which
implies that $\prove \alpha \Imp \neg (\beta \until \gamma)$.
\eprf

The following shows that the pre-model almost satisfies the truth
definitions for formulas in the basic closure. Note that every state
directly decides all formulas in the basic closure.
Define a {\em $\step$-sequence\/} of states to be a (finite or infinite)
sequence $s_1, s_2, \ldots$ such that $s_1 \step s_2 \step \ldots$.

\lem\label{lem:pre-truth}
For all $\sigma$-states $s$, we have
\begin{enumerate}
\item[(a)] 
if $\Circ \vp \in \cl_0(\psi)$, then for all states $t$
such that $s \step t$, we have
      $s\forces \Circ \vp$ iff $t\forces \vp$,
\item[(b)]
If $\ki \vp \in \cl_0(\psi)$, then $s \forces \neg \ki\vp$ iff there is
some $\sigma$-state $t$ such that $s \peqi t$ and $t \forces \neg \vp$.  Moreover, if
$|\sigma\#i| \le d$, then
$s \forces \neg \ki\vp$ iff there is
some $(\shi)$-state $t$ such that $s \peqi t$ and $t \forces \neg \vp$.

\item[(c)]
if $\vpo \until \vptw \in \cl_0(\psi)$ then $s\forces\vpo \until \vptw$
      iff there exists a $\step$-sequence $s=s_{0} \step
s_{1}\step \ldots \step s_{n}$,
      where $n\geq 0$, such that $s_{n}\forces\vptw$, and $s_{k}\forces
\vpo$ for all $k<n$.
\end{enumerate}
\elem

\prf
For part (a), suppose first that $s\forces \Circ \vp$
and $s\step t$. Since $\vp \in cl_{0}(\psi)$, it follows that
$t\forces\vp$
or $t\forces \neg \vp$. But, 
by T1 and T2, 
the latter would contradict the
assumption that $\vps\And \Circ \vptt$ is consistent. Hence we have
$t\forces \vp$.  Conversely, suppose that 
$s\step t$ and 
$t\forces \vp$.
Using T1, we have $\prove \Circ \vptt \Imp \Circ \vp$.
Since $\Circ \vp\in cl_{0}(\psi)$ we have either $s\forces\Circ\vp$ or
$s\forces \neg \Circ\vp$.
But the latter would contradict $s\step t$, so we obtain
$s\forces \Circ\vp$.

For the ``if'' direction of part (b), note that the fact that $\ki\vp$
is in $cl_{0}(\psi)$ implies that if $s\peqi t$ and $s\forces \ki\vp$,
then $t\forces \ki\vp$, hence $t\forces \vp$ by K3. For the
converse, suppose
that $t\forces \vp$ for all $\sigma$-states $t$ with $s\peqi t$.
Then $\prove \ksi\Imp \vp$, hence
$\prove \ki\ksi \Imp \ki\vp$, using K2 and R2. By
Lemma~\ref{lem:kforce}(b),
we have $s\forces \ki\ksi$. It follows immediately that $s\forces
\ki\vp$.
If $|\sigma\#i| \le d$, a similar argument shows that if $t \forces \vp$
for all $(\shi)$-states $t$ such that $s \peqi t$, then $s \forces \ki
\vp$.

For part (c), note that if $\vpo \until \vptw$ is in $cl_{0}(\psi)$,
then every state directly decides each of the formulas $\vpo$, $\vptw$,
and $\vpo\until\vptw$.
We first show that if there exists a sequence of states
$s = s_{0}\step s_{1} \step \ldots \step s_{n}$ such that
$s_{n}\forces \vptw$ and $s_{k}\forces \vpo$ for all $k<n$ then
$s\forces \vpo \until \vptw$.
We proceed by induction on $n$.  
The case $n=0$ is immediate
from T3. 
For the
general case, notice that it follows from the induction hypothesis that
$s_1 \forces (\vpo \until \vptw)$.  
Since $s_0 \step s_1$, it follows
that $\vp_{s_0} \And \Circ (\vpo \until \vptw)$ is consistent.  
By assumption, we also
have $s_0 \forces \vpo$.  Using T3, we see that $s_0 \forces \neg(\vpo \until
\vptw)$ would be a contradiction. Hence $s_0 \forces \vpo \until
\vptw$.

The converse follows immediately from Lemma~\ref{until} below.
\eprf

\lem\label{until}
If $\phi_s \land \vpo \until \vptw$ is consistent, then
there exists a
      $\step$-sequence $s=s_{0} \step
s_{1}\step \ldots \step s_{n}$,
      such that $\phi_{s_{n}} \land \vptw$ is consistent, and
$\phi_{s_{k}} \land \vpo$ is consistent for all $k<n$.
\elem
\prf
Suppose by way of contradiction that $\phi_s \land \vpo \until \vptw$ is
consistent and no appropriate $\step$-sequence exists.  Let $T$ be
the smallest set $S$ of states such that (i) $s\in S$, and (ii) if $t\in
S$, $t\step u$, and $s_u \land \vpo$ is consistent, then $u\in S$. Then
we have
that $t\forces \neg \vptw$ for all $t\in T$, so $\prove \vpT\Imp \neg
\vptw$.  In addition, for each $t\in T$ and state $u$ such that $t\step u$,
we have either $u\in T$ or $u\forces \neg \vpo\And \neg \vptw$. Thus,
using Lemma~\ref{lem:nforce}, we obtain $\prove \vpT \Imp \Circ (\vpT
\Or(\neg \vpo \And \neg \vptw))$. It now follows using
Lemma~\ref{lem:until} that $\prove \vpT \Imp \neg (\vpo \until
\vptw)$. In particular, since $s\in T$, we have $s\forces \neg
(\vpo\until \vptw)$, which contradicts the assumption that $\phi_s \land
\vpo \until \vptw$ is consistent. \eprf

For the next result, recall that when the formula $\psi$ contains
the common knowledge operator we take $d=0$, so that all states
are $\epsilon$-states.

\lem\label{Cforce}
If $C \vp \in cl_0(\psi)$, then $s\forces \neg C\vp$ iff there
is a state $t$ reachable from $s$
through the relations $\peqi$
such that $t\forces\neg\vp$.
\elem

\prf
The implication from right to left is a
straightforward consequence of the fact that if $t\forces\neg \vp$
then $t\forces \neg C\vp$, by C1, C2 and K3, together with the
fact
that if $t\peqi t'$, then $t\forces C\vp$ if and only if  $t'\forces
C\vp$.
(Proof of the latter fact: If $t \forces C\phi$ then $t \forces K_i
C\phi$ by C1 and C2.  Hence, since $t \peqi t'$
and $K_iC\vp \in cl_0(\psi)$,
we must have
$t' \forces C \phi$.  The opposite direction follows symmetrically.)
This leaves only the implication from
left to right,
for which we prove the contrapositive. Suppose that no state containing
$\neg \vp$ is reachable from $s$ by means of a sequence of steps
through the relations $\peqi$. Let $T$ be the set of states reachable
from $s$.
By Lemma~\ref{lem:kforce}(a),
if $t$ and $t'$ are
states with $t\not \peqi t'$ then $t\forces K_i\neg \vp_{t'}$. It follows
from this that $t\forces K_i \phi_T$ for every state $t \in T$ and agent
$i$. Thus, because $\prove \phi_T \Imp \vp$ we have $\prove \phi_T \Imp
E(\phi_T \And
\vp)$. By RC1 it follows that $\prove \phi_T \Imp C\vp$. Since $s\in
T$, it is immediate that $s\forces C\vp$.
\eprf

We say that an infinite $\step$-sequence of states
($s_{0},s_{1},\dots$),
where $s_{n}=(\sigma,X_{n})$ for all $n$,  is {\em acceptable\/} if
for all $n\geq 0$, if $\vpo \until \vptw\in X_{n}$ then there
exists an $m\geq n$ such that $s_{m}\forces \vptw$ and
$s_{k}\forces\vpo$ for all $k$ with $n\leq k<m$.

\dfn\label{defn:enriched-system}
An {\em enriched system}
for $\psi$
is a pair $(\runs, \Sigma)$, where $\runs$ is
a set of runs and $\Sigma$ is a {\em partial\/} function
mapping points in $\runs\times {\bf N}$ to states for $\psi$ such
that the following hold, for all runs $r\in \runs$:
\begin{enumerate}
\item If $\Sigma(r,n)$ is defined then $\Sigma(r,n')$ is defined for all
$n'>n$, and $\Sigma(r,n),\Sigma(r,n+1), \ldots$ is an acceptable
$\step$-sequence.
\item For all points $(r,n)\sim_i(r',n')$, if $\Sigma(r,n)$ is
defined then $\Sigma(r',n')$ is defined and $\Sigma(r,n) \peqi
\Sigma(r',n')$.
\item If $\Sigma(r,n)$
and $s$ are $\sigma$-states such that
$\Sigma(r,n) \peqi s$, then
there exists a point $(r',n')$ such that $(r,n) \sim_i (r',n')$
and $\Sigma(r',n') = s$.
\item if $C \phi \in cl_0(\psi)$ and $\Sigma(r,n) \forces \neg C \phi$,
then there exists a point $(r',n')$ reachable from $(r,n)$ such that
$\Sigma(r',n') \forces \neg \phi$.
\end{enumerate}

An {\em enriched$^+$ system\/} for $\psi$ is a pair $(\R,\Sigma)$
satisfying conditions 1, 2, and the following modification of 3:
\begin{itemize}
\item[3$'$.] If $\Sigma(r,n)$
is a $\sigma$-state and $s$ is a $(\shi)$-state such that
$\Sigma(r,n) \peqi s$, then
there exists a point $(r',n')$ such that $(r,n) \sim_i (r',n')$
and $\Sigma(r',n') = s$.
\end{itemize}
\edfn

Given an enriched (\respc enriched$^+$) system $(\runs, \Sigma)$, we
obtain an
interpreted system $\sysI= (\runs,\pi)$ by defining the valuation $\pi$ on
basic propositions $p$ by $\pi(r,n)(p)= {\bf true}$
just when $\Sigma(r,n)$ is defined and $\Sigma(r,n) \forces p$.
The following theorem gives a sufficient condition for a
formula in the basic closure to hold at a point in this standard
system.
If $\sigma$ is the index $i_1 \# \ldots \# i_k$,
let $K_\sigma \phi$ be an abbreviation for $K_{i_1} \ldots K_{i_k}
\phi$. (If $\sigma = \epsilon$, then we take $K_\sigma \phi$ to be
$\phi$.)

\thm\label{lem:enriched-truth}
\begin{itemize}
\item[(a)]
If $(\runs,\Sigma)$ is an enriched system for $\psi$, $\sysI$ is
the associated 
interpreted
system,
$\vp$ is in the basic
closure $cl_{0}(\psi)$, and $\Sigma(r,n)$ is defined, then
$(\sysI,r,n)\models \vp$ if and only if $\Sigma(r,n)
\forces\vp$.
\item[(b)] If $(\runs,\Sigma)$ is an enriched$^+$ system for $\psi \in
\KL_m$, $\sysI$ is the associated standard system,
$\vp$ is in the basic
closure $cl_{0}(\psi)$, $\Sigma(r,n)$ is a $\sigma$-state,
and $\ad(K_\sigma \phi) \le d$, then
$(\sysI,r,n)\models \vp$ if and only if $\Sigma(r,n)
\forces\vp$.
\end{itemize}
\ethm

\prf
We first prove part (a).
We proceed
by induction on the complexity of $\vp$. If $\vp$ is a propositional
constant then the result is immediate from the definition of $\sysI$.
The cases where $\vp$ is of the form $\neg \vpo$ or $\vpo \And \vptw$ are
similarly trivial. This leaves five cases:

{\em Case 1:\/} Suppose that $\vp$ is
of the form $\Circ \vpo$.
Then $(\sysI,r,n)\models \vp$ if and only if $(\sysI,r,n+1)\models \vpo$.
Note that $\Sigma(r,n+1)$ must be defined
by Condition 1 of Definition~\ref{defn:enriched-system}.
Since $\vpo$ is a subformula of
$\vp$ it is
in $cl_0(\psi)$,
so it follows by the induction
hypothesis that $(\sysI,r,n+1)\models \vpo$ holds precisely when
$\Sigma(r,n+1)\forces\vpo$.  By Condition 1, $\Sigma(r,n)\step
\Sigma(r,n+1)$, so we obtain from Lemma~\ref{lem:pre-truth}(a) that
$\Sigma(r,n+1)\forces\vpo$ if and only if $\Sigma(r,n)\forces\Circ
\vpo$. Putting the pieces together, we get $(\sysI,r,n)\models \vp$ if
and only if $\Sigma(r,n)\forces\vp$.

{\em Case 2:\/} Suppose that $\vp$ is
of the form $\vpo \until \vptw$.
Then the
subformulas $\vpo$ and $\vptw$ are also
in $cl_0(\psi)$.
Note
also that by Condition 1 of Definition~\ref{defn:enriched-system},
$\Sigma(r,n')$ is defined for all $n'\geq n$, and $\Sigma(r,n),
\Sigma(r,n+1), \ldots$ is an admissible $\step$-sequence. Thus, if
$\Sigma(r,n)\forces\vpo
\until \vptw$, then by Lemma~\ref{lem:pre-truth}(c)
there exists some $n'\geq n$ such that
$\Sigma(r,n')\forces\vptw$ and $\Sigma(r,k)\forces \vpo$ for $n\leq
k<n'$. By the induction hypothesis, this implies that
$(\sysI,r,n')\models \vptw$ and $(\sysI,r,k)\models \vpo$ for $n\leq
k<n'$.  In other words, we have $(\sysI,r,n)\models \vpo
\until \vptw$.  Conversely, if $(\sysI,r,n)\models \vpo \until \vptw$,
then by the induction hypothesis and the semantics of $\until$ we have
that there exists some $n'\geq n$ such that $\Sigma(r,n')\forces \vptw$
and $\Sigma(r,k)\forces \vpo$ for $n\leq k<n'$. Since $\Sigma(r,n)\step
\Sigma(r,n+1)\step \ldots \step \Sigma(r,n')$, it follows using
Lemma~\ref{lem:pre-truth}(c) that $\Sigma(r,n)\forces\vpo \until\vptw$.

{\em Case 3:\/} Suppose that $\vp$ is of the form $\ki\vpo$.
We first show
that $\Sigma(r,n)\forces\ki\vpo$ implies $(\sysI,r,n)\models \ki\vpo$.
Assume $\Sigma(r,n)\forces \ki \vpo$ and suppose that $(r,n)\eqi
(r',n')$. Then by Condition 2 of
Definition~\ref{defn:enriched-system},
we have that
that $\Sigma(r',n')$ is defined and
$\Sigma(r,n) \approx_i
\Sigma(r',n')$.
Since $\ki \vp\in cl_{0}(\psi)$ we
obtain $\Sigma(r,n')\forces \ki\vpo$. By K3 this implies
$\Sigma(r,n')\forces \vpo$.
Since $\phi \in cl_{0}(\psi)$,
by the
induction hypothesis, we obtain that $(\sysI,r',n')\models \vpo$.  This
shows that $(\sysI,r',n')\models \vpo$ for all points $(r',n')\eqi
(r,n)$.  That is, we have $(\sysI,r,n) \models \ki\vpo$.

For the converse, suppose that $\Sigma(r,n)\forces \neg\ki\vpo$
and that $\Sigma(r,n)$ is a $\sigma$-state.
By Lemma~\ref{lem:pre-truth}(b),
there exists a $\sigma$-state $t$ such that $\Sigma(r,n) \peqi t$ and
$t\forces\neg\vpo$.
By Condition 3 of
Definition~\ref{defn:enriched-system},
there exists a
point $(r',n')$ such that $(r,n)\eqi(r',n')$
and $\Sigma(r',n')=t$.
Using the induction hypothesis we obtain that
$(\sysI, r',n')\models\neg \vpo$.  It follows that $(\sysI,r,n)\models
\neg \ki\vpo$.

{\em Case 4:} If $\phi$ is of the form $E \phi_1$, the result follows
easily from the induction hypothesis, using axiom C1.

{\em Case 5:} Suppose $\phi$ is of the form $C \phi_1$.
By Condition 2 of
Definition~\ref{defn:enriched-system} we have that $\Sigma(r',n')$ is
defined for all $(r',n')$ reachable from $(r,n)$.  
An easy induction on
the length of the path from $(r,n)$ to $(r',n')$, using 
the fact that $K_iC\vpo$ is in the basic closure and axioms C1,
C2, and K3, can
be used to show that $\Sigma(r',n') \forces C \vpo$ for each point
$(r',n')$ reachable from $(r',n)$.  
Using C1, C2, and K3, it is easy to
see that $\Sigma(r',n')\forces\vpo$.  By the induction hypothesis,
this implies that
$(\sysI,r',n')\models \vpo$.  Thus, $(\sysI,r,n) \models C
\vpo$.

For the converse, suppose that $\Sigma(r,n) \forces \neg C\vpo$.
Then by Condition 4 of
Definition~\ref{defn:enriched-system}, we have $\Sigma(r',n') \forces
\neg \vpo$ for some point $(r',n')$ reachable from $(r,n)$.  By the
induction hypothesis, we have that $(\sysI,r',n') \models \neg \vpo$,
and hence $(\sysI,r,n) \models \neg C \vpo$.

For part (b), since $\psi \in \KL_m$, we only need to check the
analogues of cases 1, 2, and 3 above.  The proofs in cases 1 and 2 are
identical to those above.  The proof of case 3 is also quite similar,
but we must be a little careful in applying the inductive hypothesis.
So suppose that $\phi$ is of the form $\ki\vpo$, $\Sigma(r,n)$ is a
$\sigma$-state, and $\ad(K_\sigma \phi) \le d$.  The implication from
left to right, showing that if $\Sigma(r,n) \forces \ki\phi$ then
$(\sysI,r,n) \models \ki\phi$ is identical to that above.  We just need
the observation that if $(r,n) \sim_i (r',n')$, then $\Sigma(r',n')$ is
a $\tau$-state, where $\tau\#i = \sigma\#i$.  It follows that
$\ad(K_\tau \phi) \le \ad(K_{\sigma\#i} \phi) \le \ad(K_\sigma \ki \vp)
\le d$, so we can apply the inductive hypothesis to conclude that
$(\I,r',n') \models \phi_1$.  For the converse, the proof is again
similar.  Note that if $\ad(K_{\sigma} \ki\phi) \le d$, then
$|\sigma\#i| \le d$, so by Lemma~\ref{lem:pre-truth}(b),
there exists a $(\shi)$-state $t$ such that $\Sigma(r,n) \peqi t$ and
$t\forces\neg\vpo$.
By Condition 3$'$ of
Definition~\ref{defn:enriched-system},
there exists a
point $(r',n')$ such that $(r,n)\eqi(r',n')$
and $\Sigma(r',n')=t$.
Since $\ad(K_{\sigma\#i} \phi) = \ad(K_\sigma\ki \phi) \le d$,
using the induction hypothesis we obtain that
$(\sysI, r',n')\models\neg \vpo$.  It follows that $(\sysI,r,n)\models
\neg \ki\vpo$. \eprf

\cor\label{enriched}
If $(\runs,\Sigma)$ is an enriched (\respc enriched$^+$) system
for $\psi$, $\sysI$ is the associated 
interpreted
system, and $(r,n)$ is a point of $\sysI$ such
that $\Sigma(r,n)$
is an $\epsilon$-state and $\Sigma(r,n) \forces \psi$, then
$(\sysI,r,n) \models \psi$.
\ecor

We apply this corollary in all our completeness proofs, constructing an
appropriate enriched or enriched$^+$ system in all cases.

\section{Proofs of Soundness and Completeness}\label{proofs}

We are now in a position to prove the completeness results claimed in
Section~\ref{axioms}. Sections~\ref{proof:sync-uis}-\ref{proof:pr-sync}
will deal with the cases involving only perfect
recall, synchrony and unique initial states. The cases involving no
learning are a little more complex, and are dealt with in
Sections~\ref{proof:nl}-\ref{proof:nl-pr-sync}.

\subsection{Dealing with $\C_m$, $\C_m^{\sync}$, $\C_m^{\uis}$, and
$\C_m^{\sync,\uis}$ (Theorems~\protect{\ref{S5U}}
and~\protect{\ref{S5CU}})}
\label{proof:sync-uis}
The fact that S5C$_m^U$ is sound for $\C_m$, the class of all systems,
is straightforward and left to the reader
(see also \cite{FHMV}).
To prove completeness of S5$_m^U$ for the language $\KL_m$ and of
S5C$_m^U$ for the language $\CKL_m$ with respect to
$\C_m$, $\C_m^{\sync}$, $\C_m^{\uis}$, and
$\C_m^{\sync,\uis}$, we construct an enriched system, and use
Corollary~\ref{enriched}.
The proof proceeds in the same way whether or not common knowledge is in
the language.  We assume here that the language includes common
knowledge and that we are dealing with the axiom system S5C$_m^U$
when constructing the states in the enriched structure.
Recall that in this case we work with $\epsilon$-states only.

The following result suffices for the generation of the acceptable
sequences required for the construction of an enriched system in the
cases not involving no learning; a more complex construction will be
required in the presence of no learning.

\lem\label{acceptable}
Every finite $\step$-sequence of states
can be extended to an infinite acceptable sequence.
\elem
\prf
To see this,  first note that for every $\sigma$-state $s$ there
exists a state $t$ with $s\step t$.  For otherwise, $s\forces \neg
\Circ \vptt$ for all $\sigma$-states $t$, 
which contradicts $\prove
\Circ \bigvee_{t\ {\rm a}\; \sigma{\rm -state}} \vptt$. 
(Note that $\prove \Circ \bigvee_{t\ {\rm a}\; \sigma{\rm -state}}
\vptt$ follows from Lemma~\ref{atomiclem} and RT1.)
Thus every
finite sequence of states can be extended to an infinite sequence, and
it remains to show that the obligations arising from the
until formulas can be satisfied.

Suppose the finite $\step$-sequence is $s_0 \step \ldots \step s_n$,
where $s_k=(\sigma,X_k)$ for $k=1\ldots n$.
Now, for any formula $\vpo \until\vptw\in X_{0}$, it follows using T3
and the fact that the $s_{i}$ directly decide each of the formulas
$\vpo$, $\vptw$, and $\vpo\until \vptw$ that either the obligation
imposed by $\vpo \until \vptw$ at $s_{0}$ is already satisfied in
the sequence $(s_{0},\ldots,s_{n})$, or else $s_{n}\forces \vpo
\until \vptw$ and $s_{k}\forces \vpo$ for $0\leq k\leq n$. In the
latter case, by Lemma~\ref{lem:pre-truth}(c), there exists a sequence
$s_{n}\step s_{n+1}\step \ldots \step s_{n'}$ such that
$s_{n'}\forces\vptw$ and $s_{k}\forces\vpo$ for $n\leq k<n'$. This gives
a finite extension of the original sequence that satisfies the
obligation imposed by $\vpo \until \vptw$ at $s_{0}$. Applying this
argument to the remaining obligations at $s_{0}$, we eventually obtain
a finite sequence that satisfies all the obligations at $s_{0}$. We
may then move on to $s_{1}$ and apply the same procedure. It is clear
that in the limit we obtain an acceptable sequence extending the
original sequence.
\eprf

For each agent $i$, define the function $O_i$ to map the state
$(\sigma,U)$ to the pair $(\shi,U/\ki)$.
$O_i$ is also used
later in our other constructions.
Given a state $s$, we call $O_i(s)$ agent $i$'s {\em current
information\/} at $s$.
Let $x$ be a new object not equal to any state.  We say that a sequence
$S=(x,x,\ldots , x, s_N, s_{N+1},\ldots)$ is an {\em acceptable
sequence from $N$\/} if it starts with $N$ copies of $x$ and the suffix
$(s_N,s_{N+1},\ldots)$ is an acceptable $\step$-sequence of states for
$\psi$. Given a sequence $S$ acceptable from $N$, we define a run
$r$ as follows. For each agent $i$, take $r_i(n) = (n,S)$ when $n<N$
and $r_i(n) = (n,O_i(s_n))$ otherwise. For the environment component
$e$, take $r_e(n) = S_n$.

Let $\runs^{\sync}$ be the set of all runs so obtained, and
define the partial function $\Sigma$ on points in $\R^{\sync}\times{\bf
N}$ so that $\Sigma(r,n) = s_n$ when $r$ is derived from
a sequence
$(x,x,\ldots , x, s_N, s_{N+1},\ldots)$
acceptable from $N$ and $n\geq N$, and
$\Sigma(r,n)$ is undefined otherwise.

\lem
The pair $(\runs^{\sync},\Sigma)$ is an enriched system.
\elem

\prf
It is immediate from the construction that $(\runs^{\sync},\Sigma)$ satisfies
conditions 1 and 2 of Definition~\ref{defn:enriched-system}.  To see
that it satisfies Condition 3, suppose that $(r,n)$ is a point
such that $\Sigma(r,n)$ is defined and $\Sigma(r,n)\peqi s$. By
Lemma~\ref{acceptable} there exists an acceptable sequence $(s_n,
s_{n+1}, \ldots)$ with $s= s_n$.
Let $r'$ be the run obtained from the sequence
$(x,\ldots, x, s_n,s_{n+1}, \ldots )$.  Then it is immediate that
$(r,n)\eqi (r',n')$ and $\Sigma(r',n') = s$.
Finally, to see that it satisfies Condition 4, suppose that
$C \phi \in cl_0(\psi)$ and $\Sigma(r,n) \forces \neg C \phi$.
By Lemma~\ref{Cforce}, there
is a state $t$ reachable from $\Sigma(r,n)$
through the relations $\peqi$
such that
$t\forces\neg\vp$.
An easy inductive argument on the length of the path from $\Sigma(r,n)$
to $t$, using Condition 3, shows that there is a point $(r',n')$
reachable from $(r,n)$ through the relations $\eqi$
such
that $\Sigma(r',n') = t$.  Thus, the enriched system satisfies condition
4.  \eprf

Clearly the system $\runs^{\sync}$ is synchronous, so the interpreted
system $\I$ derived from $(\runs^{\sync},\Sigma)$ is also synchronous. 
Let $s$ be an $\epsilon$-state such that $s \forces \psi$. 
Such a state must exist because $\psi$ was assumed consistent.
By Lemma~\ref{acceptable} there exists an acceptable sequence $(s_0, s_1,
\ldots)$ with $s= s_0$. 
Let $r$ be the corresponding run in $\runs^{\sync}$.
Corollary~\ref{enriched} implies that
$(\I,r,0)\models \psi$.  This establishes the completeness of the
axiomatization S5C$^U_m$ for the language $\CKL_m$
(\respc of S5$^U_m$ for the language $\KL_m$) with respect to the
classes of systems $\C_m$ and $\C_m^\sync$.
To establish completeness of these axiomatizations for the corresponding
languages with respect to the classes of systems $\C_m^\uis$ and
$\C_m^{\sync,\uis}$, we make use of the following result, which shows
that sound and complete axiomatizations for the class of systems
satisfying some subset of the properties of perfect recall and synchrony
are also sound and complete axiomatizations for the class of systems
with the same subset of these properties, but with unique initial states
in addition. This completes the proofs of Theorem~\ref{S5U}
and~\ref{S5CU}.

\lem\label{lem:uis-construction}
Suppose $x$ is a subset of $\{\nf,\sync\}$.
If $\phi \in \CKL_m$ is satisfiable with respect to $\C_m^x$,
then it is also satisfiable with respect to $\C_m^{x,\uis}$.
\elem

\prf
Suppose $\sysI = (\runs,\pi) \in \C_m^x$.
We define a system $\sysI'$ by adding a new initial state to each run
in $\runs$.  Formally, we
define the
system $\sysI' = (\runs',\pi')$ as follows. Let $l$ be some
local state that does not occur in $\sysI$ and let $s_e$ be any state
of the environment.
For each run $r \in \runs$, let $r^+$ be the run such that
$r^+(0) = (s_e,l, \ldots, l)$ and $r^+(n+1) = r(n)$.
Let $\runs' = \{r^+: r \in \runs\}$.
The valuation $\pi'$ is given by $\pi'(r,0)(p)= {\bf false}$  and
$\pi'(r,n+1)(p)= \pi(r,n)(p)$, for $n\geq 0$ and propositions $p$.
It is clear that $\sysI'$ is a system with unique initial states.
Moreover, if $\sysI$ is synchronous, then so is $\sysI'$, and
if $\sysI$ is a system with perfect recall then so is $\sysI'$.
A straightforward induction on the construction of the
formula $\vp\in \CKL_m$ now shows that, for all points $(r,n)$ in $\I$,
we have $(\sysI,r,n)\models \vp $
iff $(\sysI',r^+,n+1)\models \vp$.
\eprf

\subsection{Dealing with $\C_m^{\nf}$ and
$\C_m^{\nf,\uis}$ (Theorem~\protect{\ref{KT3}})}

We want to show that S5$^U_m + {\rm
KT3}$ is sound and complete with respect to $\C_m^{\nf}$.   We
first consider soundness.  As we observed above, all axioms and rules of
inference other than
KT3 are known to be sound in {\em all\/} systems, so their soundness in
systems $\C_m^{\nf}$ is immediate.  The next result establishes
soundness of KT3.

\lem\label{KT3sound}
All instances of KT3 are valid in $\C_m^{\nf}$.
\elem

\prf
To show that KT3 is
sound, we assume  that $(\sysI,r,n)
\models \ki\vpo \And \Circ(\ki\vptw \And \neg \ki\vpth)$. We show that
$(\sysI,r,n) \models L_i ((\ki\vpo) \until [(\ki\vptw) \until
\neg \vpth])$.  Now it follows from the assumption that
$(\sysI,r,n+1)\models \neg \ki\vpth$, so there exists a point $(r',n')$
such that $(r,n+1)\eqi (r',n')$ and $(\sysI,r',n') \models \neg \vpth$.
Since $\I \in \C_m^{\nf}$, by condition (d) of
Lemma~\ref{nfdef}
either (i) $(r,n)\eqi (r',n')$ or (ii) there exists a number $l< n'$
such that $(r,n)\eqi (r',l)$ and
$(r,n+1)\eqi(r',k)$ for all $k$ with $l<k\leq n'$.
We claim that in either case $(\sysI,r,n)\models
L_i ((\ki\vpo) \until [(\ki\vptw) \until \neg \vpth])$.
In case (i), since $(\sysI,r',n')\models \neg
\vpth$, we have
$(\sysI,r',n')\models  (\ki\vpo) \until [(\ki\vptw) \until \neg
\vpth]$.
The desired conclusion is then immediate from the fact that
$(r,n)\eqi(r',n')$.
In case (ii), since $(\sysI,r,n)\models \ki\vpo$,
and $(r,n) \sim_i (r',l)$, we have that
$(\sysI,r',l)\models \ki\vpo$.
Similarly, because $(\sysI,r,n+1)\models \ki\vptw$, we obtain that
$(\sysI,r',k)\models \ki\vptw$ for all $k$ with $l<k\leq n'$.
Together with $(\sysI,r',n')\models \neg \vpth$, this implies that
$(\sysI,r',l)\models
(\ki\vpo) \until [(\ki\vptw) \until \neg \vpth]$.  Again, since
$(r,n)\eqi (r',l)$, we obtain that $(\sysI,r,n)\models
L_i ((\ki\vpo) \until [(\ki\vptw) \until \neg \vpth])$.
\eprf

We now establish a lemma characterizing the interaction of knowledge and
time in the pre-model. This result will enable us to satisfy the
perfect recall requirement in using the pre-model to construct an
interpreted system. It is convenient to introduce the notation $[s]_{i}$,
where $s$ is a state, for the set of $(\shi)$-states $t$ such that $s\peqi t$.
The reader is encouraged to compare the following result with
condition (d) of Lemma~\ref{nfdef}.

\lem \label{lem:trace-pr}
Suppose that the axiomatization includes \axpr.
Then for all $\sigma$-states $s,t$ and for all $(\shi)$-states $t'$, if
$s \step t$ and $t\peqi t'$, then  either (a) $s\peqi t'$ or (b)
there exists a $(\shi)$-state $s'$ such that $s\peqi s'$ and there
exists a sequence of
$(\shi)$-states
$u_{0}\step u_{1}\step \ldots \step u_{n}=t'$, where $n\geq 0$,
such that
$s'\step u_{0}$ and $u_{l} \peqi u_{l+1}$ for all $l=0 \ldots n-1$.
\elem

\prf
We derive a contradiction from the assumption that $s\step t$ and
$t\peqi t'$, but $s\not\peqi t'$ and for all $(\shi)$-states $s'$ such
that $s\peqi s'$ and all sequences of $(\shi)$-states $u_{0}\step
u_{1}\step \ldots \step u_{n}$ such that $s'\step u_{0}$ and $u_{i}
\peqi u_{i+1}$ for $i=0 \ldots n-1$, we have $u_{n} \not = t'$.
Let $T$  be the smallest set of $(\shi)$-states such that
\begin{enumerate}
\item if $v\in [s]_{i}$, $v \step v'$, and $ v'\in [t]_{i}$
then $v' \in T$, and
\item if $v\in T$, $v\step v'$, and $v'\in [t]_{i}$ then  $v'\in T$.
\end{enumerate}
Because $s\not \peqi t'$, it follows from the fact that
$\peqi$ is an equivalence relation that the intersection $[s]_{i}\cap
[t]_{i}$ is empty.  Additionally, $t'$ is not in $T$, for otherwise we
could find a sequence of the sort presumed not to exist.
Thus, for all $v \in T$, we have
$\prove \vpv
\Imp \neg \vp_{t'}$.  This implies that $\prove \vpT \Imp \neg
\vp_{t'}$.  Let $T'$ be the set of $(\shi)$-states $v'$ such
that $v\step v'$ for some $v\in T$.
We want to show that
\begin{equation}\label{eq1}
v'\forces \vpT \Or (\neg \ki\kti^+\And \neg \vp_{t'})
\end{equation}
for all $v' \in T'$.  
If $v' \in T$,
then clearly we have $v'\forces \vpT$,
so (\ref{eq1}) holds.  If $v' \notin T$, then  the second
condition in the definition of $T$ implies that $v'$ is not in
$[t]_{i}$. It follows using Lemma~\ref{lem:kforce}(d) that $v'\forces
\neg \ki \kti^+$. Further, $t\not \peqi v'$ implies that $v'\not = t'$,
so $v'\forces \neg \vp_{t'}$. Thus, again we have (\ref{eq1}).
Since (\ref{eq1}) holds for all $v' \in T$, it follows that $\prove
\vp_{T'} \Imp (\vpT \Or (\neg \ki\kti^+ \And \neg \vp_{t'}))$.  Now by
Lemma~\ref{lem:nforce}, we have $\prove \vpT \Imp \Circ \vp_{T'}$, so
using T1 and RT1 we obtain that $\prove \vpT \Imp \Circ(\vpT \Or (\neg
\ki\kti^+ \And \neg \vp_{t'}))$.  Combining this with $\prove \vpT \Imp
\neg \vp_{t'}$ and using Lemma~\ref{lem:until}, we get that
$\prove \vpT \Imp \neg (\ki\kti^+ \until \vp_{t'})$.  In particular, we
obtain $v\forces \neg (\ki\kti^+ \until \vp_{t'})$ for all states $v$ in
$T$.

We now repeat this argument to obtain a similar conclusion for the
elements of $[s]_{i}$.  Since $t'$ is not in $[s]_{i}$ we have that
$v\in [s]_{i}$ implies $v\forces \neg \vp_{t'}$.  Further, since
$[s]_{i}\cap [t]_{i}$ is empty we also have by
Lemma~\ref{lem:kforce}(d) that $v\in [s]_{i}$ implies $v\forces \neg
\ki\kti^+$. Using T3 this yields that $\prove \ksi^+ \Imp \neg (\ki\kti^+
\until \vp_{t'})$.

Let $P$ be the set of $(\shi)$-states $v'$ such that $v\step v'$ for
some $v\in [s]_{i}$.
Let $v' \in P$.  We want to show that
\begin{equation}\label{eq2}
v'\forces \ksi^+ \Or (\neg \ki\ksi^+
\And \neg (\ki\kti^+ \until \vp_{t'})).
\end{equation}
If
$v'\in [s]_{i}$ then clearly $v'\forces \ksi^+$, so (\ref{eq2}) holds.
If
$v' \notin [s]_{i}$, then, by Lemma~\ref{lem:kforce}(d), we have that
$v'\forces \neg \ki\ksi^+$.  We now consider two subcases: (a) $v'\in T$
and (b) $v'\not \in T$.  If $v'\in T$ then, as we showed earlier, we
have $v'\forces \neg (\ki\kti^+\until \vp_{t'})$.  If $v' \notin
T$, then by the definition of $T$ it follows that $t\not \peqi v'$.
By Lemma~\ref{lem:kforce}(d), this implies that $v'\forces \neg
\ki\kti^+$. Further, since $t\peqi t'$, we also obtain that $v'\not =
t'$, so $v'\forces \neg \vp_{t'}$.  Using T3, this yields
$v'\forces \neg (\ki\kti^+ \until \vp_{t'})$,
and again we have (\ref{eq2}).
Using Lemma~\ref{lem:nforce},
we obtain that $\prove \ksi^+ \Imp \Circ [\ksi^+ \Or (\neg \ki\ksi^+ \And
\neg (\ki\kti^+ \until \vp_{t'}))]$.  Applying Lemma~\ref{lem:until} to
this and the result of the preceding paragraph establishes that $\prove
\ksi^+ \Imp \neg (\ki\ksi^+ \until (\ki\kti^+ \until
\vp_{t'}))$.

It follows using Lemma~\ref{lem:kforce}(b), R2, and K2 that $s\forces
\ki\neg (\ki\ksi^+ \until (\ki\kti^+ \until \vp_{t'}))$.  By KT3, we obtain
$s\forces \neg ( \ki\ksi^+ \And \Circ (\ki\kti^+ \And L_i
\vp_{t'}))$.  Since, by Lemma~\ref{lem:kforce}(b), $s\forces \ki\ksi^+$,
we obtain using T2 that $s\forces \Circ \neg (\ki\kti^+ \And L_i
\vp_{t'}))$.  Because $s\step t$, we have that $\vps \And \Circ
\vptt$ is consistent, so it follows that $\vptt \And \neg ( \ki\kti^+
\And L_i \vp_{t'})$ is consistent. But, by
Lemma~\ref{lem:kforce}, $t\forces \ki\kti^+ \And L_i
\vp_{t'}$, so this is a contradiction.
\eprf

We are now ready to define, for each consistent $\psi$, an enriched$^+$
system
for $\psi$
that establishes completeness of ${\rm
S5}^{U}_{m}+\axpr$ with repect to $\C_m^\nf$.  The runs of this system
are those derived from the acceptable sequences $(s_0,s_1,\ldots)$
(of states for $\psi$) by putting $r_e(n) = s_n$ and $r_i(n) =
O_i(s_0)\# \ldots \# O_i(s_n)$, for each agent $i$ and $n\geq 0$.
Thus, $r_i(n)$ is the sequence of current information that agent $i$ has
had up to time $n$.
Let $\runs^{\nf}$ be the set of runs defined in this way.
The
function $\Sigma$ is given by $\Sigma(r,n)= s_n$ for each $n\geq 0$.

\lem\label{lem:enriched-pr}
Suppose that the axiomatization includes \axpr.
Then $(\runs^{\nf},\Sigma)$ is an enriched$^+$ system.
\elem

\prf
It is clear that $(\runs^{\nf},\Sigma)$ satisfies Conditions 1 and 2 of
Definition~\ref{defn:enriched-system}.
It remains to show that
Condition 3$'$ holds.  So suppose that $\Sigma(r,n)$ is a
$\sigma$-state and $\Sigma(r,n) \peqi
s$ for some $(\shi)$-state $s$.  We
must find a point $(r',n')$ such that $\Sigma(r',n') = s$.

The proof proceeds by induction on $n$. The result for $n=0$ is
immediate, since we can take $r'$ to be an acceptable
sequence starting from $s$ (such a sequence
exists by Lemma~\ref{acceptable}), so $\Sigma(r',0) = s$ and clearly
$(r,0) \sim_i (r',0)$.

Now suppose $n > 0$ and the result holds for $n-1$. Because
$\Sigma(r,n-1) \step \Sigma(r,n)$ and $\Sigma(r,n) \peqi s$,
it follows by Lemma~\ref{lem:trace-pr} that either (a)
$\Sigma(r,n-1) \peqi s$ or (b) there exists a $(\shi$)-state $s'$ such
that $\Sigma(r,n-1) \peqi
s'$ and there exists a sequence of $(\shi)$-states $u_{0}\step
u_{1}\step \ldots \step u_{k}$  such that $s'\step u_{0}$,
$u_{l} \peqi u_{l+1}$ for $l=0 \ldots k-1$, and $u_k = s$.
By the induction hypothesis, there exists for any $(\shi)$-state
$t$ with $\Sigma(r,n-1) \peqi t$ a point $(r',n')$ such that
$(r,n-1) \sim_i (r',n')$ and
$\Sigma(r',n') = t$.  In case (a),
we take $t = s$, and
we then have that $\Sigma(r,n-1) \peqi \Sigma(r,n)$ and $\Sigma(r',n') = s$.
It follows that $(r,n) \sim_i (r',n-1)$, and by the transitivity of
$\sim_i$, we also have $(r,n) \sim_i (r',n')$.  Hence we are done.
In case (b),
we take $t= s'$.
Suppose that $r'$ is derived from the sequence
$(v_{0},v_{1},\ldots)$.
Let $r''$ be any run
derived from an acceptable sequence with
initial segment $(v_{0},\ldots, v_{n'},u_{0},\ldots,u_{k})$.
Again, such a run exists by Lemma~\ref{acceptable}.  By construction,
$\Sigma(r'',n'+k+1) = u_k = s$.  Moreover, since $r''_i(n') = r'_i(n') =
r_i(n-1)$
and $O_i(u_l) = O_i(s)$ for all $l = 0, \ldots, k$,
we have $r''_i(n'+k+1) = r''_i(n')\# O_i(u_0) \# \ldots \# O_i(u_k) =
r_i(n-1) \# O_i(s) = r_i(n)$, and hence $(r,n) \sim_i (r'',n'+k+1)$.
\eprf

Now take any
$\epsilon$-state $s$ such that $s \forces \psi$, and let $r$
be a run derived from an acceptable sequence starting with $s$.  By
construction, the system $\sysI$ obtained from the enriched$^+$ system
is in $\C^\nf_m$, and by Corollary~\ref{enriched}, we have
$(\sysI,r,0)\models \psi$.
Thus, $\psi$ is satisfiable
in $\C^\nf_m$.  By Lemma~\ref{lem:uis-construction}, $\psi$ is
also satisfiable in systems in $\C^{\nf,\uis}_m$.
Since this argument applies to any formula $\psi$ consistent with
respect to ${\rm S5}_m^U+\axpr$, this
completes the proof of Theorem~\ref{KT3}.

\subsection{Dealing with $\C_m^{\nf,\sync}$ and $\C_m^{\nf,\sync,\uis}$
(Theorem~\protect{\ref{KT2}})}
\label{proof:pr-sync}
We now show that ${\rm S5}_m^U+\axprsync$ is sound and complete
with respect to  $\KL_m$ for the classes of systems $\C^{\nf,\sync}_m$
and $\C^{\nf,\sync,\uis}_m$. For soundness, the following result suffices.

\lem\label{KT2sound}
All instances of \axprsync\ are valid in $\C^{\nf,\sync}_m$. \elem

\prf
Let $\sysI$ be a system in $\C^{\nf,\sync}_m$ and let $r$ be a run
of $\sysI$. Suppose that $(\sysI,r,n)\models K_i\Circ\vp$.
If $(r,n+1) \sim_i (r',n')$, then by synchrony we must have $n'=n+1$.
Thus, by perfect recall and synchrony, we have $(r,n)\sim_i (r',n'-1)$.
It follows that $(\sysI,r',n'-1) \models \Circ \vp$, which implies that
$(\sysI,r',n') \models \vp$. This shows that
$(\sysI,r',n') \models \vp$ for all $(r',n')\sim_i(r,n+1)$.
Thus, we have $(\sysI,r,n+1) \models K_i \vp$, and hence
$(\sysI,r,n) \models \Circ K_i \vp$.
\eprf

Before constructing an enriched$^+$ system for the completeness part, we
first note a property of the pre-model, analogous to
Lemma~\ref{lem:trace-pr}.

\lem \label{lem:trace-pr-sync}
Suppose that the axiomatization includes \axprsync.  Then for all
$\sigma$-states $s,t$ with $s \step t$, we have that for all
$(\shi)$-states $t'$ with $t\peqi t'$ there exists a
$(\shi)$-state $s'$ such that $s\peqi s'$ and $s' \step t'$.
\elem

\prf
By way of contradiction, suppose that $s,t$ are $\sigma$-states with
$s\step t$
that $t'$ is a $(\shi)$-state such that $t \peqi t'$, but that for all
$(\shi)$-states $s'$ such that $s \peqi s'$, we have that  $s'\forces
\neg \Circ \phi_{t'}$.  By T2, we have that $s' \forces \Circ \neg
\phi_{t'}$ for all $(\shi)$-states $s'$ such that $s \peqi s'$.  By
Lemma~\ref{lem:kforce}(b),  it follows that $s \forces K_i \Circ
\neg \phi_{t'}$.
By \axprsync, we have that $s \forces \Circ K_i \neg \phi_{t'}$.
Since $s \step t$, it follows that $\phi_t \land K_i \neg \phi_{t'}$
is consistent.  However, since $t \peqi t'$, by
Lemma~\ref{lem:kforce}(c), we have $t \forces L_i \phi_{t'}$.
This is a contradiction.
\eprf

To construct the enriched$^+$ system, we now
take $\runs^{\nf,\sync}$ to be the set of runs $r$ derived from
acceptable sequences
$(s_0,s_1,\ldots)$ of states for the formula $\psi$ by putting $r_e(n)
= s_n$ and $r_i(n) = O_i(s_0)\ldots O_i(s_n)$, for each agent $i$ and $n\geq
0$.
The notation $O_i(s_0) \ldots O_i(s_n)$ is meant to denote the
sequence formed by concatening $O_i(s_0), O_i(s_1), \ldots, O_i(s_n)$.
Thus, the length of the sequence is $n+1$, which enforces synchrony.
Again, the function $\Sigma$ is given by $\Sigma(r,n)= s_n$ for each
$n\geq 0$.

\lem\label{lem:enriched-prsync}
Suppose that the axiomatization includes \axprsync.
Then $(\runs^{\nf,\sync},\Sigma)$ is an enriched$^+$ system.
\elem

\prf
Conditions 1 and 2 of the definition of an enriched$^+$ system are
immediate.
To show that Condition 3$'$ holds, suppose
that $\Sigma(r,n)$ is a $\sigma$-state, and $t$ is a $(\shi)$-state such
that $\Sigma(r,n) \peqi t$.  Suppose that $r$ is derived from the acceptable
sequence $(s_0,s_1, \ldots)$, so $\Sigma(r,n)=s_n$.  It follows from
Lemma~\ref{lem:trace-pr-sync} that there exists
a $\step$-sequence $t_0\step \ldots \step t_n$ such that $t_n=t$ and
$s_j\peqi t_j$ for $j=1\ldots n$.  By Lemma~\ref{acceptable}, this
sequence may be extended to an infinite acceptable sequence. Taking
$r'$ to be the run derived from this sequence, we see that
$(r,n) \sim_i (r',n)$ and $\Sigma(r',n)= t$.
\eprf

Take $\sysI^{\nf,\sync}$ to be the system obtained from
$(\runs^{\nf,\sync},\Sigma)$.
By construction, this system is in $\C^{\nf,\sync}$.
Now take any
$\epsilon$-state $s$ such that $s \forces \psi$, and let $r$ be a
run derived from an acceptable sequence starting with $s$.  By
construction, the system $\sysI$ obtained from the enriched$^+$ system
is in $\C^{\nf,\sync}_m$, and by Corollary~\ref{enriched}, we have
$(\sysI,r,0)\models \psi$.
Thus, $\psi$ is satisfiable
in $\C^{\nf,\sync}_m$.  By Lemma~\ref{lem:uis-construction}, $\psi$ is
also satisfiable in systems in $\C^{\nf,\sync,\uis}_m$.
Since this argument applies to any formula $\psi$ consistent with
respect to ${\rm S5}_m^U+\axprsync$, this
completes the proof of Theorem~\ref{KT2}.

\subsection{Dealing with $\C_m^{\nl}$ (Theorem~\protect{\ref{KT4}})}
\label{proof:nl}

We want to show that ${\rm S5}^U_m + {\rm KT4}$ is sound and complete
for $\KL_m$ with respect to $\C_m^{\nl}$.  For soundness, it suffices to
show that KT4 is valid in $\C_m^{\nl}$.  This is straightforward.

\lem\label{KT4sound} All instances of KT4 are valid in $\C_m^{\nl}$.
\elem
\prf Suppose that $\I \in \C_m^{\nl}$ and $(\I,r,n) \sat K_i \phi \until
K_i \psi$.  We want to show that $(\I,r,n) \sat K_i(K_i \phi \until K_i
\psi)$.  Thus, if
$(r',n') \eqi (r,n)$, we must show that  $(\I,r',n') \sat K_i \phi
\until K_i \psi$.  Since $(\I,r,n) \sat K_i \phi \until K_i \psi$, there
exists $l \ge n$ such that
$(\I,r,l) \sat K_i \psi$ and $(\I,r,k) \sat
K_i \phi \land \neg K_i \psi$ for all $k$ with $n \le k < l$.
Note that this means that if $n \le k < l$, then $r_i(k) \ne r_i(l)$.
Since $\I \in \C_m^{\nl}$ and $(r,n) \eqi (r',n')$, there must be some
$l' \ge n'$
such that $((r,n), \ldots, (r,l))$ is $\sim_i$-concordant with $((r,n'),
\ldots, (r,l'))$.
Thus, there exists some $h$, a
partition $S_1, \ldots, S_h$ of the sequence
$((r,n), \ldots, (r,l))$, and a partition $T_1, \ldots, T_h$ of the
sequence $((r',n'), \ldots, (r,l'))$ such that for all $j = 1, \ldots, h$,
we have $(r,k) \sim_i (r',k')$ for all points $(r,k) \in S_j$ and
$(r',k') \in T_j$.  It easily follows that $(\I,r',n') \sat
K_i \phi \until K_i \psi$, as desired.
\eprf

For completeness, we define an appropriate enriched$^+$ system.
As we shall see,
the demands of no learning make this a little more subtle than in the
case of no forgetting.

For the remainder of this section, consistency and provability are with
respect to
a logic that includes
${\rm S5}^U_m + {\rm KT4}$.
Fix a consistent formula $\psi$
such that $\ad(\psi) = d$.
Our first step is to prove an analogue of Lemma~\ref{lem:trace-pr}.
\lem\label{extend}
Suppose that the axiomatization includes KT4.
If $s$ is a $\sigma$-state, $t$ is a $(\shi)$-state, $s \peqi t$,
and $s \step s'$, then there exists a sequence $t_0, \ldots, t_k$ such
that (a) $t = t_0$, (b) $t_j \peqi s$ for $j < k$, (c) $t_j \step
t_{j+1}$ for $j < k$, and (d) $s' \peqi t_k$. \elem

\prf
If $s' \peqi t$, then we can take the sequence to consist only of $t$,
and we are done.  Otherwise,
since $\phi_s \land \Circ \phi_{s'}$ is consistent,
it follows from Lemma~\ref{lem:kforce}(b) that $\phi_s \land
K_i \Phi^+_{s,i} \until K_i \Phi^+_{s',i}$ is consistent.  Moreover, by
Lemma~\ref{lem:kforce}(c), we have that $\phi_s \forces L_i
\phi_t$. Thus, $\phi_s \land L_i \phi_t \land
K_i \Phi^+_{s,i} \until K_i \Phi^+_{s',i}$ is consistent.  Using KT4, it
follows that $\phi_t \land
K_i \Phi^+_{s,i} \until K_i \Phi^+_{s',i}$ is consistent.
The result now follows from Lemma~\ref{until}.
\eprf

Unfortunately, Lemma~\ref{extend} does not suffice to construct
an enriched$^+$ system.  Roughly speaking, the problem is the following.
In the case of perfect recall, we used
Lemma~\ref{lem:trace-pr} to show that, given a $\step$-sequence
$S = (s_0, \ldots, s_n)$ of
$\sigma$-states and a $(\shi)$-state $t$ such that
$s_n \peqi t$, we can construct a $\step$-sequence $T$ of
$(\shi)$-states
ending  with $t$ such that $S$ is $\peqi$-concordant with $T$.
There is
no problem then extending $T$ to an acceptable sequence.
Moreover, we can extend $S$ and $T$ independently to acceptable sequences;
all that matters is that the finite
prefixes of these sequences---namely,
$S$ and $T$---are $\peqi$-concordant.  With no learning, on the other
hand, it is the infinite suffixes that must be $\peqi$-concordant.
Given a $\step$-sequence $S = (s_0, \ldots)$ of $\sigma$-states
and a $(\shi)$-state $t$ such that $s_0 \peqi t$, using
Lemma~\ref{extend}, we can find a
$\step$-sequence $T$ starting with $t$ that is
$\peqi$-concordant with $S$.
This suggests that it is possible to find the appropriate
sequences for the construction of runs satisfying the no learning condition.
Unfortunately, it does not follow from the acceptability of
$S$ that $T$ is also acceptable. This makes it necessary to work with
a smaller set of sequences than the set of all acceptable sequences,
and to build up the sequences $S$ and $T$ simultaneously.
To ensure that the appropriate obligations are satisfied at all
points in the set of runs constructed, we need
to work not just with single states, but with trees of states.

A {\em $k$-tree for $\psi$\/} (with $k \le d$)
is a set $S$ of
$\sigma$-states for $\psi$ with $|\sigma| \le k$ with a unique
$\epsilon$-state
such that
if $s \in S$ is a $\sigma$-state then
\begin{itemize}
\item if $t$ is a $(\shi)$-state
such that $s \peqi t$ and $|\sigma\#i| \le k$, then $t \in S$,
\item if $\sigma = \tau \# i$, then
there is a $\tau$-state $t$ in $S$ such that $s \peqi t$.
\end{itemize}
We extend the $\step$ relation to $k$-trees as follows.  If $S_1$ and
$S_2$ are $k$-trees for $\psi$, then $S_1 \step_f S_2$ if
$f$ is a function associating with each $\sigma$-state $s \in S_1$ a
finite sequence of $\sigma$-states in $S_1 \union S_2$
such that
\begin{itemize}
\item if $f(s) = (s_0, \ldots, s_k)$, then
\begin{itemize}
\item $s = s_0$,
\item $s_0 \step \cdots  \step s_k$,
\item $s_0, \ldots, s_{k-1} \in S_1$ and $s_k \in S_2$;
\end{itemize}
\item if $s \peqi s'$, then $f(s)$ and $f(s')$ are $\peqi$-concordant;
\item for at least one $s \in S_1$, the sequence $f(s)$ has length at
least 2.
\end{itemize}

Given two sequences of $\sigma$-states $\alpha = (s_0, \ldots, s_k)$ and
$\beta = (t_0, \ldots)$, where
$\alpha$ is finite, the {\em fusion\/} of $\alpha$ and $\beta$, denoted
$\alpha \cdot \beta$ is defined only if $s_k = t_0$; in this case, it is
the sequence $(s_0, s_{k-1}, t_0, \ldots)$.  Given an infinite sequence
${\cal S} = S_{0} \step_{f_0} S_1 \step_{f_1} S_2 \step_{f_2} \ldots$
of $k$-trees,
we say a sequence $\alpha$ of $\sigma$-states is
{\em compatible with ${\cal S}$\/} if there exists some $h$, and
$\sigma$-states $s_h$, $s_{h+1}, \ldots$ with $s_j \in S_j$ for $j \ge
h$,  such that $\alpha = f_h(s_h) \cdot f_{h+1}(s_{h+1}) \cdot \ldots$.
(Implicit in this notation is the assumption that this fusion product is
defined, so that the last state in $f_j(s_j)$ is the same as the first
state in $f_{j+1}(s_{j+1})$, for $j \ge h$.)
A $\step$-sequence $(t_0,t_1,\ldots)$ of $\sigma$-states is
a {\em compression\/} of $(s_0, s_1, \ldots)$ if (1) $t_0 = s_0$, and
(b) if $t_j = s_h$, then $t_{j+1}$ is $s_{h'}$, where $h'$ is the least
integer greater than $h$ such that $s_{h'-1} \step
s_{h'}$ and $s_h = \cdots = s_{h'-1}$.  (If no such $h'$ exists, then
$t_j$ is the last element of the
compression.)
${\cal S}$ is {\em acceptable} if every
$\step$-sequence that is a compression of some
sequence compatible with ${\cal S}$
is infinite and acceptable. Our goal
is to construct an acceptable sequence of $d$-trees; we shall use this
to define the enriched$^+$ system.

Note that by Lemma~\ref{lem:kforce}, the formula $\vp_s$ essentially
describes the subtree below $s$ of any $k$-tree containing $s$.
Given a $k$-tree $S$ and a $\sigma$-state $s$ in $S$, we inductively
define a formula
$\treep_{S,s}$ that describes all of $S$ from
the point of view of $s$.  If $s$ is an $\epsilon$-state, then
$\treep_{S,s} = \vp_s$.  Otherwise, if $s$ is a
($\tau\#i$)-state, where $\tau \ne \tau\#i$, then $$\treep_{S,s} =
\vp_s \land \bigand_{\{\tau {\rm -states}\;  t:\; s \peqi t\}} L_i
\treep_{S,t}.$$

If $S$ and $T$ are $k$-trees, $s \in S$, and $t \in T$, then we write
$(S,s) \step^+
(T,t)$ if there exists a sequence of $k$-trees $S_0, \ldots, S_l$ and
functions $f_0, \ldots, f_{l-1}$ such that $S_0 \step_{f_0} \cdots
\step_{f_{l-1}} S_l$, $S_0 = S$, $S_l = T$,  $f_j(s) = (s)$ for $j
\le l-2$, and $f_{l-1}(s) = (s,t)$.

\lem\label{path}
Suppose that the axiomatization includes KT4, $S$ is a $k$-tree, and $s$
is a $\sigma$-state in $S$, where $|\sigma| = k$.
\begin{itemize}
\item[(a)]  If
$t$ is a $\sigma$-state and
$\treep_{S,s} \land \Circ
(\phi_t \land \xi)$ is consistent,
then there
exists a $k$-tree $T$ such that $t \in T$, $(S,s) \step^+ (T,t)$, and
$\treep_{T,t} \land \xi$ is consistent.
\item[(b)]
$\treep_{S,s} \rimp \Circ \bigor_{\{(T,t):\;
(S,s) \step^+ (T,t)\}} \treep_{T,t}$ is
provable.
\item[(c)] If $\treep_{S,s} \land \phi \until \phi'$ is consistent, then
for some $l\ge 0$
there is
a sequence $S_0, \ldots, S_l$ of $k$-trees and states $s_0, \ldots, s_l$
such that (i) $s_j \in S_j$, (ii) $(S,s) = (S_0, s_0)$, (iii)
$(S_j,s_j) \step^+ (S_{j+1},s_{j+1})$ for $j = 0, \ldots, l-1$,
(iv) $\treep_{S_j,s_j} \land \phi$ is consistent for $j = 0, \ldots,
l-1$, and (v) $\treep_{S_l,s_l} \land \phi'$ is consistent.
\end{itemize}
\elem

\prf
We proceed by induction on $k$.
The case that $k=0$ is immediate using standard arguments, since then
$\treep_{S,s}$ is just $\phi_s$.

So suppose $k > 0$ and $\sigma = \tau\#i$, with $\sigma \ne \tau$.
We first prove part (a) in the case that $\xi$ is of the form $K_i
\xi'$, then part (b), then prove the general case of (a), and then prove
(c).  First consider (a) in the case that $\xi$ is of the form $K_i
\xi'$.  Note that $\treep_{S,s} \land \Circ
(\phi_t \land K_i\xi')$ implies
$\treep_{S,s} \land K_i\Phi_{s,i}
\until K_i(\xi' \land \Phi_{t,i})$.
{F}rom the definition
of $k$-tree, it follows that there is a $\tau$-state $s'$ in $S$ such
that $s \peqi s'$.  Let
$S'$ be the $(k-1)$-tree
consisting of all $\sigma'$-states in $S$ with $|\sigma'| \le k-1$.
{F}rom KT4, it follows that
$$\treep_{S',s'} \land K_i\Phi_{s,i}
\until K_{i}(\xi' \land \Phi_{t,i})$$
is consistent.  Applying part (c) of the inductive hypothesis, we get a
sequence $S_0, \ldots, S_l$ of ($k-1$)-trees
and states $s_0, \ldots, s_l$
such that (i) $s_j \in S_j$, (ii) $(S',s') = (S_0, s_0)$, (iii)
$(S_j,s_j) \step^+ (S_{j+1},s_{j+1})$ for $j = 0, \ldots, l-1$,
(iv) $\treep_{S_j,s_j} \land K_i\Phi_{s,i}$ is consistent for $j = 0,
\ldots, l-1$, and (v) $\treep_{S_l,s_l} \land K_i(\xi' \land
\Phi_{t,i})$ is consistent.
It follows by definition that
there is a sequence $T_0, \ldots, T_m$ of $(k-1)$-trees
and functions $f_0, \ldots, f_{m-1}$ such that
$T_0 \step_{f_0} \step \cdots \step_{f_{m-1}} T_m$, $T_0 = S_0$,
and $T_m = S_l$.  Moreover, there are elements $t_0, \ldots, t_m$
such that $t_0 = s'$, $t_m = s_l$, if $j < m$, then $t_j = s_{j'}$ for
some $j' \le j$,
and if $t_j = t_{j+1}$, then
$f_j(t_j) = (t_j)$, while if $t_j \ne t_{j+1}$, then
$f_j(t_j) = (t_j, t_{j+1})$, for $j = 0, \ldots, m-1$.

Let $T_j'$ be the unique $k$-tree extending $T_j$, for $j = 0, \ldots,
m$.  Since $\phi_{t_j} \land K_i \Phi_{s,i}$ is consistent for $j < m$,
we have that
$t_j \peqi s$, and so $s \in T_j'$ for $j < m$.  Similarly, we have
that $t \in T_m'$.  We now show how to construct $f_j'$, for $j < m$.
For each state $u' \in T_j' - T_j$, there must exist a
state $u \in T_j$ and an agent $j'$ such that $u \approx_{j'} u'$.
(There may be more than one such state $u$, of course.
In this case, in the construction below, we can pick $u$
arbitrarily.)
It easily follows from Lemma~\ref{extend} that there exists a
sequence $\alpha_{u'}$ starting with $u'$
that is $\approx_{j'}$-concordant
with $f_j(u)$.
Moreover, we can take $\alpha_{t_j} = (s)$ for $j < m-1$,
and take $\alpha_{t_{m-1}} = (s,t)$.
We define $f_j'$ so that it agrees with $f_j$ on $T_j$, and for
each $u' \in T_j'-T_j$, we have $f_j'(u') = \alpha_{u'}$.

Notice that $T_0' = S$.  If $m > 0$, it follows immediately from the
definition that $(S,s) \step^+ (T_m,t)$, and that $\treep_{T_m,t} \land
K_i \xi'$ is consistent.  If $m=0$, it is easy to check that
we must have $t \in S$, for we have $s' \peqi t$.  Since we also
have $s' \peqi s$, it follows that $s \peqi t$.  Define $f$ so that
$f(u) = (u)$ for $u \ne s$ and $f(s) = (s,t)$.  Then $(S,s) \step_f
(S,t)$.  Since $s \step t$, we have $(S,s) \step^+ (S,t)$.  This
completes the proof of part~(a).

To prove part (b), suppose not.  Then
$ \treep_{S,s}   \land
    \Circ \bigand_{\{ (T,t): \, (S,s) \step^+ (T,t)\}}
                     \neg \treep_{T,t}$ is
consistent.
Straightforward temporal reasoning shows that there must be
some $u$ such that
\begin{equation}\label{eqnl1}
\treep_{S,s}   \land
    \Circ(\phi_u \land  \bigand_{\{ (T,t): \, (S,s) \step^+ (T,t)\}}
                     \neg \treep_{T,t})
\end{equation}
is consistent.  Now $\neg \treep_{T,t}$ is equivalent to $\neg \phi_t
\lor
\bigor_{\{\tau {\rm -states}\ t':\; t' \peqi t\}} K_i \neg
\treep_{T,t'}$.  Thus, it follows that the consistency of (\ref{eqnl1})
implies that for each tree $T$ such that $(S,s) \step^+ (T,u)$, there is
a $\tau$-state
$t_T \peqi u$ such that
\begin{equation}\label{eqnl2}
\treep_{S,s}   \land
    \Circ(\phi_u \land  K_i
(\bigand_{\{ T: \, (S,s) \step^+ (T,u)\}}
\neg \treep_{T,t_T}))
\end{equation}
is consistent.  By part~(a), there
exists a $k$-tree $T'$ and $t' \in T'$ such that $(S,s) \step^+
(T',t')$ and $\treep_{T',t'} \land \phi_u \land
K_i (\bigand_{\{ T: \, (S,s) \step^+ (T,u)\}} \neg \treep_{T,t_T})
$ is consistent.
But this means that $t' = u$.  Thus, we have a contradiction, since
$\treep_{T',u} \land
K_i \neg \treep_{T',t_{T'}}$ is inconsistent.

The general case of (a) follows easily from (b).
Part (c) also follows from part (b),
using arguments much like those of Lemma~\ref{until};
we omit details here.
\eprf

\lem\label{acceptable1} Suppose that the axiomatization includes KT4
and $\psi$ is consistent.
Then there is an acceptable sequence of $d$-trees such
that $\psi$ is true at the root of the first tree.
\elem

\prf The key part of the proof is to show that given a finite sequence
$S_0 \step_{f_0} \ldots \step_{f_{l-1}} S_{l}$ of $d$-trees and a
$\sigma$-state $s$ in
$S_l$ such that $s \forces \Circ \phi$ (\respc $s \forces \phi_1
\until \phi_2$), we can extend the sequence of trees in such a way as to
satisfy this obligation.  This follows easily from
Lemmas~\ref{extend} and~\ref{path}.  In more detail, suppose
$s \forces \phi_1 \until \phi_2$.  Let $S'$ be consist of all
$\tau$-states in $S_l$, with $|\tau| \le k = |\sigma|$.
By Lemma~\ref{path}, we can
find a sequence of $k$-trees starting with $S'$ that satisfies
this obligation.  Using Lemma~\ref{extend}, we can extend this to a
sequence of $d$-trees starting with $S_l$ that satisfies the obligation.
The argument in the case that $s_k \forces \Circ \phi$ is similar.
We can then take care of the obligations one by one, and construct an
acceptable sequence, in the obvious way.

Since $\psi$ is consistent, there must be some tree $S$ with root $s_0$
such that  $s_0 \forces \psi$.  We just extend $S$ as above to complete
the proof. \eprf

Once we have an acceptable sequence ${\cal S}$ of $d$-trees as in
Lemma~\ref{acceptable1}, we can easily construct
the enriched$^+$ system much as we did in the case of perfect recall.
Given a $\step$-sequence $s_0 \step s_1 \step \ldots$
the {\em $\nl$-run\/} $r$
derived from it is defined so that
$r_e(n) = s_n$ and $r_i(n) =
O_i(s_n)\# O_i(s_{n+1}) \# \ldots$, for each agent $i$ and $n\geq 0$.
Thus, while for perfect recall, we take $r_i(n)$ to consist of the
agent's current information up to time $n$, for no learning, we take
$r_i(n)$ to consist of the current information from time $n$ on.
The construction stresses the duality between perfect recall and no
learning.
Let $\runs^{\nl}$ consist of all the $\nl$-runs derived from
$\step$-sequences that are compressions of sequences of states
compatible with ${\cal S}$.   Again,
the function $\Sigma$ is given by $\Sigma(r,n)= s_n$ for each $n\geq 0$.

\lem\label{nlsystem} Suppose the axiomatization includes KT4.  Then
$(\runs^{\nl}, \Sigma)$
is an enriched$^+$ system.
\elem
\prf Again, Conditions 1 and 2 in the definition of enriched$^+$ system
follow immediately from the construction.  For Condition 3$'$, suppose
that
$(r,n)$ is a point in the system, $\Sigma(r,n)$ is a $\sigma$-state, and
$s$ is a $(\shi)$-state
with $s \peqi \Sigma(r,n)$.
Suppose ${\cal S} = S_0 \step_{f_0} S_1 \step_{f_1} S_2 \step_{f_2}
\ldots$. By definition, the run
$\nl$-run
$r$ is derived from the compression of
some sequence $(s_0, s_1, s_2, \ldots)$
of $\sigma$-states compatible with ${\cal S}$.
Suppose $\Sigma(r,n)$ is in the interval of this sequence from
$S_k$. Then $s$ must also be in $S_k$.
Let $(t_0, t_1, \ldots)$ be the (unique) sequence compatible with
${\cal S}$ that starts at $s$ in $S_k$.
Let $r'$ be
the run derived from the compression of this
sequence.  Then, by
definition, we have $\Sigma(r',0) = s$ and $(r,n) \sim_i (r',0)$.
\eprf

Take $\sysI^\nl$ to be the system obtained from
$(\runs^{\nl},\Sigma)$.
By construction, this system is in $\C^{\nl}$.
Now take any
$\epsilon$-state $s$ such that $s \forces \psi$, and let $r$ be a
run derived from the compression of
a sequence compatible with $\cal S$ starting with $s$.
It follows that $(\sysI,r,0) \sat \psi$.
Thus, $\psi$ is satisfiable
in $\C^\nl_m$.  This
completes the proof of Theorem~\ref{KT4}.

\subsection{Dealing with $\C_m^{\nl,\nf}$ and
$\C_1^{\nl,\nf,\uis}$ (Theorem~\protect{\ref{KT3+4}})}
We now want to show that ${\rm S5}^U_m + {\rm KT3} + {\rm KT4}$ is sound
and complete for $\KL_m$ with respect to $\C_m^{\nl,\nf}$.
Soundness is immediate from Lemmas~\ref{KT3sound} and~\ref{KT4sound}.

For completeness, we construct an enriched$^+$ system much as in the
proof of Theorem~\ref{KT4}, using $k$-trees.
By Lemma~\ref{acceptable1}, there is an acceptable sequence $\cal S$ of
$d$-trees such that $\psi$ is true at the root of the first tree.
Given a $\step$-sequence $(s_0,s_1,....)$,
the {\em nl-nf run\/} $r$
derived from it is defined so that
$r_e(n) = s_n$ and $r_i(n) = (O_i(s_0)\#O_i(s_1) \# \ldots \# O_i(s_n),
O_i(s_n)\# O_i(s_{n+1}) \# \ldots)$, for each agent $i$ and $n\geq 0$.
Thus, the agents' local states enforce both perfect recall (by keeping
track of all the information up to time $n$) and no learning (by keeping
track of the current information from time $n$ on).
Let $\runs^{\nl,\nf}$ consist of all nl-nf-runs that are derived
from $\step$-sequences that have a suffix that is
the compression of a sequence of states compatible with $\cal S$.
Note that now we consider $\step$-sequences whose {\em suffixes\/} are
compressions of sequences compatible with ${\cal S}$.  The reason to
allow the greater generality of suffixes will become clear shortly.
Since $\cal S$ is acceptable, it is easy to see that every such
$\step$-sequence must be infinite and acceptable.
 Again,
the function $\Sigma$ is given by $\Sigma(r,n)= s_n$ for each $n\geq 0$.

\lem\label{nfnlsystem} Suppose the axiomatization includes KT3 and KT4.
Then $(\runs^{\nl,\nf}, \Sigma)$
is an enriched$^+$ system.
\elem

\prf  As usual, Conditions 1 and 2 in the definition of enriched$^+$
system follow immediately from the construction.  For Condition 3$'$,
suppose that
$(r,n)$ is a point in the system, $\Sigma(r,n)$ is a $\sigma$-state, and
$s$ is a $(\shi)$-state.
Suppose ${\cal S} = S_0 \step_{f_0} S_1 \step_{f_1} S_2 \step_{f_2}
\ldots$. By definition, the run $\nl$--$\nf$-run $r$ is derived from a
$\step$-sequence $(s_0, s_1, \ldots)$ that has a suffix $(s_N,
s_{N+1}, \ldots)$ that is the compression of some sequence $(t_0,
t_1, t_2, \ldots)$
of $\sigma$-states compatible with ${\cal S}$,
and $\Sigma(r,n) = s_n$.
There are now two cases to consider.
If $n \ge N$, then there exists some $k$ such that $s_n$ is in $S_k$.
Then $s$ must also be in $S_k$.
Let $(u_0,u_1, \ldots)$ be the unique sequence compatible with ${\cal S}$
that starts with $s$ in $S_k$.
By Lemma~\ref{lem:trace-pr},
there exist $\shi$-states $v_0 \step \cdots \step v_h$
such that $v_h = s$ and $(v_0, \ldots, v_h)$ is $\peqi$-compatible with
$(s_0, \ldots, s_n)$.  Consider the $\step$-sequence formed from the
fusion
of $(v_0, \ldots, v_h)$ and the compression of $(u_0, u_1, \ldots)$.  By
construction, the
$\nl$--$\nf$-run $r'$ derived from this $\step$-sequence is in
$\runs^{\nl,\nf}$, $\Sigma(r',h) = s$, and $(r,n) \sim_i (r',h)$.

Now suppose
$n<N$.
By Lemma~\ref{lem:trace-pr}, there exist
$(\shi)$-states $v_0 \step \cdots \step v_h$ such that $v_h = s$ and
$(v_0, \ldots, v_h)$ is
$\peqi$-concordant with $(s_0,...,s_n)$. Moreover, by
Lemma~\ref{extend},
this sequence can be extended to a sequence
$(v_0,....v_k)$ that is $\peqi$-concordant with $(s_0,...,s_N)$. Since
$s_N \in S_M$ for some $M$, we must have $v_k \in S_M$.
Let $(u_0, u_1, \ldots)$ be the unique sequence compatible with
${\cal S}$ that starts at $v_k$ in $S_M$.
Consider the $\step$-sequence formed from the
fusion
of $(v_0, \ldots, v_k)$ and the compression of $(u_0, u_1, \ldots)$.
By construction, the
$\nl$--$\nf$-run $r'$ derived from this $\step$-sequence is in
$\runs^{\nl,\nf}$, $\Sigma(r',h) = s$, and $(r,n) \sim_i (r',h)$. \eprf

Again, we complete the proof by taking $\sysI^{\nl,\nf}$ to
be the system obtained from $(\runs^{\nl,\nf},\Sigma)$.
By construction, this system is in $\C^{\nl,\nf}$ and satisfies $\psi$.
This shows that
${\rm S5}^{U}_{m}+{\rm KT3 + KT4}$ is a sound and
complete axiomatization for the language $\KL_{m}$ with respect to
$\C_m^{\nl,\nf}$ for all $m$.

The fact that it is also a sound and complete
axiomatization for the language $\KL_{1}$ with respect to
$\C_1^{\nl,\nf,\uis}$ follows immediately from the following lemma.

\lem\label{adduis}  The formula $\psi \in \KL_1$ is satisfiable with
respect to $\C_1^{\nl,\nf}$ (\respc $\C_1^{nl,\nf,\sync}$)
iff it is satisfiable with respect to $\C_1^{\nl,\nf,\uis}$ (\respc
$\C_1^{\nl,\nf,\sync,\uis}$). \elem

\prf Clearly if $\psi$ is satisfiable with respect to
$\C_1^{\nl,\nf,\uis}$ (\respc $\C_1^{\nl,\nf,\sync,\uis}$)
it is satisfiable with respect to
$\C_1^{\nl,\nf}$ (\respc $\C_1^{\nl,\nf,\sync}$).
For the converse, suppose that $(\I,r^*,n^*) \sat \psi$,
where $\I = (\R,\pi) \in \C_1^{\nl,\nf}$.
For each run $r \in \R$, define the run $r^+$ just as in
Lemma~\ref{lem:uis-construction}, to be the result of adding a new
initial state to $r$. Let $\R' = \{r^+: (r,0) \sim_1 (r^*,0)\}$.
Define $\pi'$ as on $\R'$ as in Lemma~\ref{lem:uis-construction}, so
that $\pi'(r^+,0)(p) = {\bf false}$ for all primitive propositions
$p$, and $\pi'(r^+,n+1) = \pi(r,n)$.  Let $\I' = (\R',\pi')$.
Clearly $\I' \in \C_1^{\nl,\nf,\uis}$, and if $\I$ is synchronous,
then so is $\I'$.
 We claim that $(\I',r^+,n+1) \sat \phi$ iff $(\I,r,n) \sat \phi$ for
 all formulas $\phi \in \KL_1$ and all $r^+ \in \R'$.
We prove this
 by induction on the structure of $\phi$.  The only nontrivial
case is if $\phi$ is of the form $K_1 \phi'$.  But this case is
immediate from the observation that if
$r^+ \in \R'$,
$r' \in \R$, and $(r,n) \sim_1 (r',n')$, then since $\I$ is a
system of perfect recall, we must have $(r,0) \sim_1 (r',0)$, and
hence $(r')^+ \in \R'$.
We leave details of the proof of the claim to the
reader.  {F}rom the claim, it follows that $\psi$ is satisfiable in
$\C_1^{\nl,\nf,\uis}$, and that if $\psi$ is satisfiable in
$\C_1^{\nl,\nf,\sync}$, then it is also satisfiable in
$\C_1^{\nl,\nf,\sync,\uis}$.
\eprf

\subsection{Dealing with $\C_m^{\nl,\sync}$
(Theorem~\protect{\ref{KT5}})}

We now want to show that ${\rm S5}^U_m + {\rm KT5}$ is sound
and complete for $\KL_m$ with respect to $\C_m^{\nl,\sync}$.
Soundness follows from the following lemma.

\lem\label{KT5sound}
All instances of KT5 are valid in $\C_m^{\nl,\sync}$.
\elem
\prf Suppose $\I \in \C_m^{\nl,\sync}$ and $(\I,r,n) \sat \Circ \ki
\vp$.  We want to show that $(\I,r,n) \sat \ki \Circ \vp$.  Thus,
suppose that $(r',n') \sim_i (r,n)$.  We must show that $(\I,r',n') \sat
\Circ \vp$.   By synchrony, we must have $n' =
n$.  Moreover, by no learning and synchrony, we have that $(r,n+1)
\sim_i (r',n+1)$.  Since $(\I,r,n+1) \sat \ki \vp$, it follows that
$(\I,r',n+1) \sat \vp$, and hence that $(\I,r',n) \sat \Circ \phi$, as
desired. \eprf

For completeness, we construct an enriched$^+$ system much as in the
proof of Theorem~\ref{KT4}, using $k$-trees, with an appropriate
strengthening of the $\step$ relation.

We start by proving the following analogue
of Lemma~\ref{extend}.
\lem\label{extendsync}
Suppose that the axiomatization includes KT5.
If $s$ is a $\sigma$-state, $t$ is a $(\shi)$-state, $s \peqi t$,
and $s \step s'$, then there exists a $(\shi)$-state $t'$ such that $t
\step t'$ and $s' \peqi t'$. \elem

\prf
Since $\phi_s \land \Circ \phi_{s'}$ is consistent,
it follows from Lemma~\ref{lem:kforce}(b) that $\phi_{s_0} \land \Circ
K_i \Phi^+_{s',i}$ is consistent.  Moreover, by
Lemma~\ref{lem:kforce}(c), we have that $\phi_s \forces L_i
\phi_t$. Thus, $\phi_s \land L_i \phi_t \land
\Circ K_i \Phi^+_{s',i}$ is consistent.  Using KT5, it
follows that $\phi_s \land L_i \phi_t \land
K_i \Circ \Phi^+_{s',i}$ is consistent.  It follows that $\phi_t \land
\Circ \Phi^+_{s',i}$ is consistent.  Hence, there is some
$(\shi)$-state $t'$
such that $s' \peqi t'$ and $\phi_t \land \Circ \phi_{t'}$ is
consistent.  Thus, we have $s' \peqi t'$ and $t \step t'$.  \eprf

If $S$ and $T$ are $k$-trees, $s \in S$, and $t \in T$, we define
$(S,s) \step^{\sync,+} (T,t)$ if
$S \step_f T$ for some $f$
such that $f(s) = (s,t)$ and $f(s')$ has length 2 for all $s' \in S$.
We now get the following simplification of Lemma~\ref{path}.

\lem\label{pathsync}
Suppose that the axiomatization includes KT5, $S$ is a $k$-tree,
and $s$ is a $\sigma$-state in $S$, where $|\sigma| = k$.
\begin{itemize}
\item[(a)] If $\treep_{S,s} \land \Circ
(\phi_t \land \xi)$ is consistent, then there
exists a $k$-tree $T$ and $t \in T$ such that $(S,s) \step^{\sync,+}
(T,t)$ and $\treep_{T,t} \land \xi$ is consistent.
\item[(b)]
$\treep_{S,s} \rimp \Circ \bigor_{\{(T,t):\; (S,s) \step^{\sync,+}
(T,t)\}}
\treep_{T,t}$ is provable.
\item[(c)] If $\treep_{S,s} \land \phi \until \phi'$ is consistent, then
there is
a sequence $S_0, \ldots, S_l$ of $k$-trees and states $s_0, \ldots, s_l$
such that (i) $s_j \in S_j$, (ii) $(S,s) = (S_0, s_0)$, (iii)
$(S_j,s_j) \step^{\sync,+} (S_{j+1},s_{j+1})$ for $j = 0, \ldots, l-1$,
(iv) $\treep_{S_j,s_j} \land \phi$ is consistent for $j = 0, \ldots,
l-1$, and (v) $\treep_{S_l,s_l} \land \phi'$ is consistent.
\end{itemize}
\elem
\prf
The proof is like that of Lemma~\ref{path}, using
Lemma~\ref{extendsync} instead of Lemma~\ref{extend}.  We leave details
to the reader. \eprf

We can then define a $\sync$-acceptable sequence
of trees by replacing $\step^+$ by $\step^{\sync,+}$ in
the definition of acceptable sequence of trees.  Using
Lemma~\ref{pathsync},
we can show that if the axiom system contains KT5, then we can
construct an infinite $\sync$-acceptable sequence ${\cal S}
= S_0\step^{\sync,+} S_1\step^{\sync,+} S_2\step^{\sync,+}\ldots$
of $d$-trees.
As in Section~\ref{proof:sync-uis}, we use an object $x$ not equal to any state.
Given a $\step$-sequence $s_N \step s_{N+1} \step \ldots$ starting at
$s_N\in S_N$
the {\em $\nl$--$\sync$-run\/} $r$
derived from it is defined so that
$r_e(n) = s_n$ for $n\geq N$, else $r_e(n)=x$, and
for each agent $i$, if $n\geq N$ then
$r_i(n) = (n, O_i(s_n)  O_i(s_{n+1})  \ldots)$,
else $r_i(n) = (n, x^{N-n} O_i(s_n)  O_i(s_{n+1})
\ldots)$.
Thus, the local state of the agent enforces synchrony (by
encoding the time) and enforces no learning.
Let $\runs^{\nl,\sync}$ consist of all $\nl$--$\sync$-runs derived from
$\step$-sequences compatible with $\cal S$,
and define
$\Sigma$ by taking $\Sigma(r,n)= s_n$
for $n\geq N$ and $\Sigma(r,n)$ undefined for $n<N$.

\lem\label{syncnlsystem} Suppose the axiomatization includes KT5.
Then $(\runs^{\nl,\sync}, \Sigma)$
is an enriched$^+$ system.
\elem
\prf The proof is essentially the same as that of Lemma~\ref{nlsystem}.
We leave details to the reader. \eprf

We complete the proof of Theorem~\ref{KT5} just as we did all the
previous proofs.

\subsection{Dealing with $\C_m^{\nl,\nf,\sync}$
(Theorem~\protect{\ref{KT2+5}})}
\label{proof:nl-pr-sync}
We now want to show that ${\rm S5}^U_m + {\rm KT2} + {\rm KT5}$ is sound
and complete for $\KL_m$ with respect to $\C_m^{\nl,\nf,\sync}$.
Soundness follows from Lemmas~\ref{KT2sound} and~\ref{KT5sound}.

For completeness, we construct an enriched$^+$ system by combining the
ideas of the proofs of Theorems~\ref{KT3+4} and~\ref{KT5}.
Using Lemma~\ref{pathsync},
we can show that if the axiom system contains KT5, then we can
construct an infinite $\sync$-acceptable sequence ${\cal S}$ of
$d$-trees.
Given a $\step$-sequence $s_0 \step s_1 \step \ldots$,
the {\em $\nl$--$\nf$--$\sync$-run\/} $r$
derived from it is defined so that
$r_e(n) = s_n$ and $r_i(n) = (O_i(s_1)   \ldots  O_i(s_n),
O_i(s_n)  O_i(s_{n+1})  \ldots)$.
Thus, the local state of the agent enforces both no forgetting
and no learning.
It also enforces synchrony, since the agent can determine $n$ from the
length of the first of the two seqeunces in its local state.
Let $\runs^{\nl,\nf,\sync}$ consist of all $\nl$--$\nf$--$\sync$-runs
derived from $\step$-sequences with suffixes that are compatible with
$\cal S$; again, we define
$\Sigma$ by taking $\Sigma(r,n)= s_n$.

Using ideas similar to those in earlier proofs, we can now prove the
following result.
\lem\label{syncnfnlsystem} Suppose the axiomatization includes KT2 and
KT5. Then
$(\runs^{\nl,\nf,\sync}, \Sigma)$
is an enriched$^+$ system.
\elem
We complete the proof of Theorem~\ref{KT2+5} just as we did the earlier
proofs.

\subsection{Dealing with $\C_m^{\nl,\sync,\uis}$ and
$\C_m^{\nl,\nf,\sync,\uis}$ (Theorem~\protect{\ref{KT2+6}})}
Finally, we want to show that
${\rm S5}^U_m + {\rm KT2} + {\rm KT5} + \{K_i \phi \equiv K_1 \phi\}$ is
sound
and complete for $\KL_m$ with respect to $\C_m^{\nl,\sync,\uis}$ and
$\C_m^{\nl,\nf,\sync,\uis}$.  Soundness follows easily using the
following result,
which is Proposition 3.9 in \cite{HV2} (restated using our notation).
\pro\label{nlsyncuis}
\begin{enumerate}
\item[(a)] $\C_m^{\nl,\sync,\uis} = C_m^{\nl,\nf,\sync,\uis}$.
\item[(b)] Any formula $\phi$ in $\KL_m$ is equivalent in
 $\C_m^{\nl,\sync,\uis}$ to the formula $\phi'$ that results
by replacing all occurrences of $K_i$, $i \ge 2$, by $K_1$.
\end{enumerate}
\epro
It follows from part~(a) of Proposition~\ref{nlsyncuis} that the same
axioms characterize $\C_m^{\nl,\sync,\uis}$ and
$\C_m^{\nl,\nf,\sync,\uis}$.  Now using Theorems~\ref{KT2sound}
and~\ref{KT5sound}, the soundness of KT2 and KT5 follows.  The soundness
of $K_i \phi \equiv K_1 \phi$ follows from part~(b).

For completeness, using the axiom $K_i \phi \equiv K_1 \phi$, it
suffices to show the completeness of
${\rm S5}^U_1 + {\rm KT2} + {\rm KT5}$ with respect to
$\C_1^{\nl,\pr,\sync,\uis}$.  By Theorem~\ref{KT2+5}, this
axiomatization is complete with respect to $\C_1^{\nl,\pr,\sync}$.  The
result now follows using Lemma~\ref{adduis}.

\section{Remarks on No Learning}\label{nolearningremarks}

We noted in Section~\ref{background} that the definition of no learning
adopted in this paper differs from that used in \cite{HV,HV2}.
We now comment on the reason for this change and the relationship
between these alternative definitions
of no learning.

First, recall
from
part (d) of
Lemma~\ref{nfdef} that
that agent $i$ has perfect recall in system $\R$  if and only if
\bi
\item[($*$)]
for all points $(r,n) \eqi (r',n')$ in $\R$, if $k\leq n$,
then there exists $k'\leq n'$ such that $(r,k) \eqi(r',k')$.
\ei
Intuitively, no learning is the dual of perfect recall, so it
seems reasonable to define no learning by replacing references to the
past in a definition of perfect recall by references to the future.  This
was done in \cite{HV,HV2}, where the definition given for no learning
was the following future time variant of condition ($*$),
which we call no learning$'$, to distinguish it from our current
definition:
Agent $i$ does not learn$'$ in system $\R$ if and only if
\bi
\item[($**$)] for all points $(r,n) \eqi (r',n')$ in $\R$, if $k\geq n$
then there exists $k'\geq n'$ such that $(r,k) \eqi(r',k')$.
\ei
The following lemma states a number of relations holding between
condition ($**$) and the other properties we have considered in this
paper.
\begin{lemma}
\be
\item[(a)] If agent $i$ does not learn in system $\R$ then
agent $i$ does not learn$'$ in system $\R$.

\item[(b)] If system $\R$ is synchronous
or if agent $i$ has perfect recall in $\R$,
then agent $i$ does not learn in $\R$ iff agent $i$ does not learn$'$ in
$\R$.
\ee
\end{lemma}
\prf
We first show (a). Suppose that agent $i$ does not learn in $\R$.
Assume that $(r,n) \eqi (r',n')$ and let $k\geq n$.
Since $i$ does not learn, the future local state sequences at
$(r,n)$ and $(r',n')$ are equal. It follows that
there exists $k'\geq n'$ such that $(r,k) \eqi (r',k')$.
Thus, agent $i$ does not learn$'$.

For (b), it follows from part (a) that
it suffices to show the implication from no learning$'$ to
no learning. We consider the cases of synchrony and perfect
recall independently. In each case, we show that if $(r,n) \eqi
(r',n')$ then there exists $k\geq n'$ such that the sequences
$((r,n),(r,n+1))$ and $((r',n') \ldots (r',k))$ are $\eqi$-concordant.
It then follows by Lemma~\ref{nldef} that agent $i$ does not learn.

Assume first that $\R$ is a synchronous system, and that
$(r,n)\eqi(r',n')$. By synchrony, we must have $n=n'$.
By no learning$'$, there exists $k\geq n$ such that
$(r,n+1)\eqi(r',k)$. By synchrony, $k$ must equal $n+1$.
It is immediate that $((r,n),(r,n+1))$ and $((r',n'),(r',n'+1))$
are $\eqi$-concordant.

Next, assume that agent $i$ has perfect recall in $\R$, and
that $(r,n)\eqi(r',n')$. By no learning$'$, there exists $k\geq n'$
such that $(r,n+1)\eqi(r',k)$. By perfect recall, agent $i$'s local
state sequences  $(r,n+1)$ and $(r',k)$ are identical, as are
the local state sequences at $(r,n)$ and $(r',n')$.
It follows that the sequences $((r,n),(r,n+1))$ and
$((r',n'), \ldots,(r',k))$ are $\eqi$-concordant.
\eprf

Thus, in the context of either synchrony or perfect recall,
no learning and no learning$'$ are equivalent.
However, in systems without
synchrony or perfect recall,
no learning$'$ is strictly weaker than no learning, as the
following example shows.
Consider the system $\R = \{r^1,r^2\}$ for a single agent, where
the runs are defined by:
\[ r^1(n)  = \left\{\begin{array}{ll}
                (s_e, a) & \mbox{if $n=0$} \\
                (s_e, b) & \mbox{if $n > 0$ is odd } \\
                (s_e, c) & \mbox{if $n>0$ is even }
              \end{array} \right. \]
where $s_e$ is some state of the environment, and $a,b,c$ are
local states of agent 1, and similarly
\[ r^2(n)  = \left\{ \begin{array}{ll}
                (s_e, a) & \mbox{if $n=0$} \\
                (s_e, c) & \mbox{if $n > 0$ is odd } \\
                (s_e, d) & \mbox{if $n>0$ is even }
                \end{array} \right.
\]
This system clearly satisfies $\uis$ and condition ($**$),
so we have no learning$'$ (for both agents).  However,
agent 1's future local state
sequences from the points $(r^1,0) \sim_1 (r^2,0)$ are not
$\sim_1$-concordant,
so we do not have no learning.
Thus, no learning and no learning$'$ are distinct in general.

This raises the question of which of variant to take as the definition
of no learning for the cases $\C^\nl$ and $\C^{\uis,\nl}$.  The origin
of this notion in the literature lies in Ladner and Reif's paper
\cite{LR}, where it is motivated as arising in the context of blindfold
games.
Their logic
LLP assumes perfect recall, so is not decisive on the distinction.
However, it seems that the behavior in the above example is somewhat
unnatural for this application, and the definition we have adopted in
this paper better fits the intuition of a player in a blindfold game
following a fixed linear strategy, but with some uncertainty about timing.
It is such examples that in fact led us to use the current definition of
no learning.

It is worth noting that the example above also shows that the axiom
\axnl\
is not sound with respect to the class of systems satisfying
($**$). Define the interpretation $\pi$ of the propositions $p$ and $q$
on runs $r \in \R$ by $\pi(r,n)(p) = {\bf true}$ iff $r_1(n) = a$ and
$\pi(r,n)(q) = {\bf true}$ iff $r_1(n) = b$. Let $\I = (\R,\pi)$. It
is then readily seen that $(\I,r^1,0) \models K_1 p \until K_1 q$ but
not $(\I,r^1,0) \models K_1(K_1 p \until K_1 q)$.
Hence \axnl\ fails in this system.
(This example is a future time version of an example used in
\cite{Mey93} to show that the axiom KT1 is incomplete for systems with
perfect recall.)
We have not investigated the issue of axiomatization using no
learning$'$ rather than no learning in the two cases where there is a
difference---$\C_m^{nl}$ and $\C_1^{nl,uis}$.  We conjecture that, while
there will be a relatively clean complete axiomatization in these cases,
it will not be as elegant as the one proposed here.  That is, the axiom
that captures no learning will be somewhat more complicated than KT4.
This conjecture is in line with our feeling that no learning is the
``right'' definition, not no learning$'$.

We remark that the complexity results of \cite{HV,HV2} are proved in
the context of no learning$'$, but it is relatively straightforward to
show that the same results hold if we use the definition of no learning
instead. 

\section{Discussion}\label{discussion}

While we have looked in this paper at the effect on axiomatization of
some combinations of classes of systems and language (48 in all!),
there are certainly other cases of interest.  One issue we have
already mentioned is that of braching time versus linear time. Basing
the temporal fragment of the language on branching time yields another
48 logics, whose complexity is studied in \cite{HV}.  We would
conjecture that the obvious translations of the axioms we have
presented here deal with branching time, with similar proofs of
completeness, but this remains to be verified.

It is worth remarking that our results are very sensitive to the
language studied.  As we have seen, the language considered in this
paper is too coarse to reflect some properties of systems.  In the
absence of the other properties, synchrony and unique initial states
do not require additional axioms.  This may no longer be true for
richer languages.  For example, if we allow past-time operators
\cite{LPZ}, we 
need not only the additional axioms capturing the
properties of these, but also new axioms describing the interaction of
knowledge and time.  Suppose that we add an operator $\ominus$ such
that $(\I,r,n) \models \ominus \vp$ if $n\geq 1$ and $(\I,r,n-1)
\models \vp$.  
Notice that $\neg \ominus \true$ expresses the property ``the time is
0'' and $\ominus \neg \ominus \true$ expresses the property ``the time is
1''.  Similarly, we can inductively define formulas that express the
property ``the time is $m$'' for each $m \ge 0$.
If {\it time=m}
is an abbreviation for this formula, then ${\it time=m} \Imp K_i({\it
time=m})$ is valid in $\C^{\sync}$, for each time m.

On the other hand, by adding past time operators we can simplify the axiom for 
perfect recall. Introducing the operator $S$ for ``since'', we may show that 
the formula
\[ (K_i \vp)S (K_i\psi) \Imp K_i ((K_i \vp)S (K_i\psi))\]
is valid in $\C^{\pr}$.  This axiom very neatly expresses the meaning
of perfect recall, and a comparison with KT4 shows clearly the sense
in which perfect recall is a dual of no learning. Techniques 
similar to those developed in this paper may be used to 
prove that this axiom, together with the usual axioms for past time
\cite{LPZ} and for knowledge, yields a complete axiomatization 
for $\C^{\pr}$. 

Besides changes to the language, there are also additional properties
of systems worth considering. One case of interest is the class of
{\em asynchronous message passing systems\/} of \cite{FHMV}.  That
extra axioms are required in such systems is known (\cite{FHMV}
Exercise 8.8), but the question of complete axiomatization is still
open.

\bibliographystyle{alpha}
\bibliography{z}
\end{document}

%% file: defn.tex
\newtheorem{THEOREM}{Theorem}[section]
\newenvironment{theorem}{\begin{THEOREM} \hspace{-.85em} {\bf :} }%
                        {\end{THEOREM}}
\newtheorem{LEMMA}[THEOREM]{Lemma}
\newenvironment{lemma}{\begin{LEMMA} \hspace{-.85em} {\bf :} }%
                      {\end{LEMMA}}
\newtheorem{COROLLARY}[THEOREM]{Corollary}
\newenvironment{corollary}{\begin{COROLLARY} \hspace{-.85em} {\bf :} }%
                          {\end{COROLLARY}}
\newtheorem{PROPOSITION}[THEOREM]{Proposition}
\newenvironment{proposition}{\begin{PROPOSITION} \hspace{-.85em} {\bf :} }%
                            {\end{PROPOSITION}}
\newtheorem{DEFINITION}[THEOREM]{Definition}
\newenvironment{definition}{\begin{DEFINITION} \hspace{-.85em} {\bf :} \rm}%
                            {\end{DEFINITION}}
\newtheorem{CLAIM}[THEOREM]{Claim}
\newenvironment{claim}{\begin{CLAIM} \hspace{-.85em} {\bf :} \rm}%
                            {\end{CLAIM}}
\newtheorem{EXAMPLE}[THEOREM]{Example}
\newenvironment{example}{\begin{EXAMPLE} \hspace{-.85em} {\bf :} \rm}%
                            {\end{EXAMPLE}}
\newtheorem{REMARK}[THEOREM]{Remark}
\newenvironment{remark}{\begin{REMARK} \hspace{-.85em} {\bf :} \rm}%
                            {\end{REMARK}}

\newcommand{\thm}{\begin{theorem}}
\newcommand{\lem}{\begin{lemma}}
\newcommand{\pro}{\begin{proposition}}
\newcommand{\dfn}{\begin{definition}}
\newcommand{\rem}{\begin{remark}}
\newcommand{\xam}{\begin{example}}
\newcommand{\cor}{\begin{corollary}}
\newcommand{\prf}{\noindent{\bf Proof:} }
\newcommand{\ethm}{\end{theorem}}
\newcommand{\elem}{\end{lemma}}
\newcommand{\epro}{\end{proposition}}
\newcommand{\edfn}{\bbox\end{definition}}
\newcommand{\erem}{\bbox\end{remark}}
\newcommand{\exam}{\bbox\end{example}}
\newcommand{\ecor}{\end{corollary}}
\newcommand{\eprf}{\bbox\vspace{0.1in}}
\newcommand{\beqn}{\begin{equation}}
\newcommand{\eeqn}{\end{equation}}

\newcommand{\bbox}{\vrule height7pt width4pt depth1pt}

\newcommand{\clm}{\begin{claim}}
\newcommand{\eclm}{\end{claim}}

\newcommand{\sat}{\models}

\newcommand{\rimp}{\Rightarrow}

\newcommand{\dimp}{\Leftrightarrow}

\newcommand{\union}{\cup}

\newcommand{\IN}{\mbox{$I\!\!N$}}

\renewcommand{\phi}{\varphi}
\newcommand{\Circ}{\mbox{{\small $\bigcirc$}}}

\newcommand{\C}{{\cal C}}

\newcommand{\I}{{\cal I}}

\newcommand{\R}{{\cal R}}

\newcommand{\<}{\langle}
\renewcommand{\>}{\rangle}

\newtheorem{cl}{Claim}

\newcommand{\ie}{i.e.,~}

\newcommand{\respc}{resp.,\ }

\newcommand{\ol}{\setlength{\itemsep}{0pt}\begin{enumerate}}
\newcommand{\eol}{\end{enumerate}\setlength{\itemsep}{-\parsep}}
\newcommand{\ul}{\setlength{\itemsep}{0pt}\begin{itemize}}
\newcommand{\dl}{\setlength{\itemsep}{0pt}\begin{description}}
\newcommand{\edl}{\end{description}\setlength{\itemsep}{-\parsep}}
\newcommand{\eul}{\end{itemize}\setlength{\itemsep}{-\parsep}}

\newcommand{\true}{\mbox{{\it true}}}

\newcommand{\untill}{U}
\newcommand{\until}{\, U \,}

\newcommand{\commentout}[1]{}

\newcommand{\bi}{\begin{itemize}}
\newcommand{\ei}{\end{itemize}}
\newcommand{\be}{\begin{enumerate}}
\newcommand{\ee}{\end{enumerate}}

%% file: hmvcorr.bbl
\begin{thebibliography}{FHMV95}

\bibitem[FHMV95]{FHMV}
R.~Fagin, J.~Y. Halpern, Y.~Moses, and M.~Y. Vardi.
\newblock {\em Reasoning about Knowledge}.
\newblock MIT Press, Cambridge, Mass., 1995.

\bibitem[FHV91]{FHV1}
R.~Fagin, J.~Y. Halpern, and M.~Y. Vardi.
\newblock A model-theoretic analysis of knowledge.
\newblock {\em Journal of the ACM}, 91(2):382--428, 1991.
\newblock A preliminary version appeared in {\em Proc.~25th IEEE Symposium on
  Foundations of Computer Science}, 1984.

\bibitem[GPSS80]{GPSS}
D.~Gabbay, A.~Pnueli, S.~Shelah, and J.~Stavi.
\newblock On the temporal analysis of fairness.
\newblock In {\em Proc.~7th ACM Symp.~on Principles of Programming Languages},
  pages 163--173, 1980.

\bibitem[Hin62]{Hi1}
J.~Hintikka.
\newblock {\em Knowledge and Belief}.
\newblock Cornell University Press, Ithaca, N.Y., 1962.

\bibitem[HM92]{HM2}
J.~Y. Halpern and Y.~Moses.
\newblock A guide to completeness and complexity for modal logics of knowledge
  and belief.
\newblock {\em Artificial Intelligence}, 54:319--379, 1992.

\bibitem[HV86]{HV}
J.~Y. Halpern and M.~Y. Vardi.
\newblock The complexity of reasoning about knowledge and time.
\newblock In {\em Proc.~18th ACM Symp.~on Theory of Computing}, pages 304--315,
  1986.

\bibitem[HV88a]{HV4}
J.~Y. Halpern and M.~Y. Vardi.
\newblock The complexity of reasoning about knowledge and time in asynchronous
  systems.
\newblock In {\em Proc.~20th ACM Symp.~on Theory of Computing}, pages 53--65,
  1988.

\bibitem[HV88b]{HV3}
J.~Y. Halpern and M.~Y. Vardi.
\newblock The complexity of reasoning about knowledge and time: synchronous
  systems.
\newblock Research Report RJ 6097, IBM, 1988.

\bibitem[HV89]{HV2}
J.~Y. Halpern and M.~Y. Vardi.
\newblock The complexity of reasoning about knowledge and time, {I}: lower
  bounds.
\newblock {\em Journal of Computer and System Sciences}, 38(1):195--237, 1989.

\bibitem[Leh84]{Leh}
D.~Lehmann.
\newblock Knowledge, common knowledge, and related puzzles.
\newblock In {\em Proc.~3rd ACM Symp.~on Principles of Distributed Computing},
  pages 62--67, 1984.

\bibitem[LPZ85]{LPZ}
O.~Lichtenstein, A.~Pnueli, and L.~Zuck.
\newblock The glory of the past.
\newblock In Rohit Parikh, editor, {\em Proc.~Workshop on Logics of Programs},
  Lecture Notes in Computer Science, Vol. 193, pages 196--218. Springer-Verlag,
  Berlin/New York, 1985.

\bibitem[LR86]{LR}
R.~E. Ladner and J.~H. Reif.
\newblock The logic of distributed protocols (preliminary report).
\newblock In J.~Y. Halpern, editor, {\em Theoretical Aspects of Reasoning about
  Knowledge: Proc.~1986 Conference}, pages 207--222. Morgan Kaufmann, San
  Francisco, Calif., 1986.

\bibitem[Mey94]{Mey93}
R.~{van}~{der} Meyden.
\newblock Axioms for knowledge and time in distributed systems with perfect
  recall.
\newblock In {\em Proc.~9th IEEE Symp.~on Logic in Computer Science}, pages
  448--457. 1994.

\bibitem[PR85]{PR}
R.~Parikh and R.~Ramanujam.
\newblock Distributed processing and the logic of knowledge.
\newblock In R.~Parikh, editor, {\em Proc.~Workshop on Logics of Programs},
  pages 256--268, 1985.

\bibitem[Sat77]{Sat}
M.~Sato.
\newblock A study of {K}ripke-style methods for some modal logics by
  {G}entzen's sequential method.
\newblock {\em Publications Research Institute for Mathematical Sciences, Kyoto
  University}, 13(2):381--468, 1977.

\bibitem[Spa90]{Spaan}
E.~Spaan.
\newblock Nexttime is not necessary.
\newblock In R.~J. Parikh, editor, {\em Theoretical Aspects of Reasoning about
  Knowledge: Proc.~Third Conference}, pages 241--256. Morgan Kaufmann, San
  Francisco, Calif., 1990.

\end{thebibliography}
